\documentclass[aps,prd,showpacs]{revtex4}

\usepackage{amsmath,amssymb,amsfonts,amsthm,bbm} 
\usepackage{color}
\usepackage{graphicx}
\usepackage{pstricks,pstricks-add}
\usepackage{pst-plot}
\usepackage[hang,nooneline]{subfigure}
\usepackage{slashbox}

\definecolor{dark-green}{rgb}{0,0.7,0}
\definecolor{dark-blue}{rgb}{0,0.2,0.5}
\definecolor{med-blue}{rgb}{0,0.7,1}
\definecolor{mblue}{rgb}{0,0.2,1}
\definecolor{cnc}{rgb}{0.8,0,0}
\definecolor{light-red}{rgb}{1,0.8,0.8}
\definecolor{dark-yellow}{rgb}{1,0.8,0}
\definecolor{light-blue}{rgb}{0.8,0.9,1}
\definecolor{grey}{rgb}{0.211,0.211,0.211}
\definecolor{verylight-blue}{rgb}{0.93,0.95,1}
\definecolor{light-yellow}{rgb}{1,0.9,0.8}

\begin{document}

\title{Test particle motion in the space--time of a Kerr black hole pierced by a cosmic string}

\author{Eva Hackmann $^{(a)}$}
\email{hackmann@zarm.uni-bremen.de}
\author{Betti Hartmann $^{(b)}$ }
\email{b.hartmann@jacobs-university.de}
\author{Claus L{\"a}mmerzahl $^{(a), (c)}$}
\email{laemmerzahl@zarm.uni-bremen.de}
\author{Parinya Sirimachan $^{(b)}$}
\email{p.sirimachan@jacobs-university.de}

\affiliation{
$(a)$ ZARM, Universit\"at Bremen, Am Fallturm, 28359 Bremen, Germany\\
$(b)$ School of Engineering and Science, Jacobs University Bremen, 28759 Bremen, Germany\\
$(c)$ Institut f\"ur Physik, Universit\"at Oldenburg, 26111 Oldenburg, Germany}
\date\today

\begin{abstract}

We study the geodesic equation in the space-time of a Kerr black hole pierced by an infinitely thin cosmic string and give the complete set of analytical solutions of this equation for massive and massless
particles in terms of Mino time that allows to decouple the
$r$- and $\theta$-component of the geodesic equation. The solutions of the geodesic equation can be classified according to the
particle's energy and angular momentum, the mass and angular
momentum per mass of the black hole. We give examples of orbits showing the influence of the
cosmic string. We also discuss the perihelion shift and the Lense-Thirring
effect for bound orbits and show that the presence of a cosmic string enhances both effects. 
Comparing our results with experimental data from the LAGEOS satellites
we find an upper bound on the energy per unit length of a string piercing the earth which is approximately
$10^{16}$ kg/m.
Our work has also applications to the recently suggested explanation of the alignment of the polarization vector
of quasars using remnants of cosmic string decay in the form of primordial magnetic field loops.

\end{abstract}

\pacs{04.20.Jb, 02.30.Hq}

\maketitle

\section{Introduction}
The motion of test particles (both massive and massless) provides the only experimentally feasible way to study the gravitational fields of objects such as black holes.
Predictions about observable effects (light--like deflection, gravitational
time--delay, perihelion shift and Lense--Thirring effect) can be
made and compared with observations. Geodesics in black hole space-times in 4 
dimensional Schwarzschild space--time \cite{hagihara} and Kerr
and Kerr--Newman space--time \cite{chandra}
have been discussed extensively.
This has been extended to the cases of Schwarzschild--de Sitter space-times \cite{hl1} as well as to spherically symmetric higher dimensional
space--times \cite{hackmann08}. 
The analytical solutions of the geodesic equation in the Kerr space--time have been presented in \cite{chandra,ONeill}.
Moreover, solutions of the geodesic equation in the Kerr space--time have been given using 
elliptic functions \cite{Kraniotis2004}, while 
solutions representing bound orbits parameterized in terms of {\it Mino time} \cite{Mino2003} have been presented in \cite{Fujita2009}. Spherical orbits in the Kerr--(Anti)--de Sitter space--time have  been discussed in \cite{Kraniotis2004}, while the general solution
to the geodesic equation in (4--dimensional) Kerr--de Sitter \cite{hackmannetal1} and even general Plebanski--Demianski space--times
without acceleration has been found \cite{hackmannetal2}.

In \cite{DrascoHughes2004} a Fourier expansion has been used to compute the fundamental
frequencies for bound orbits using Mino time. These results have direct application
in the computation of gravitational waves that are created in extreme mass ratio inspirals, i.e. in binaries in which
a stellar object moves on a bound orbit around a supermassive black hole.

Cosmic strings have gained a lot of renewed interest over the past years
due to their possible connection to string theory \cite{polchinski}.
These are topological defects \cite{vs} that could have formed in one of the numerous phase transitions in the early universe due to the Kibble mechanism.
Inflationary models resulting from string theory (e.g. brane inflation)
predict the formation of cosmic string networks at the end of inflation \cite{braneinflation}.

Different space-times containing cosmic strings have been discussed in the past.
This study has mainly been motivated by the pioneering work of Bach and Weyl \cite{bw} 
describing a pair of black holes held apart by an infinitely thin strut.
This solution has later been reinterpreted in terms of cosmic strings describing
a pair of black holes held apart by two cosmic strings extending to infinity in opposite direction.
Consequently, a cosmic string piercing a Schwarzschild black hole has also been discussed, both in the thin string limit \cite{afv} -- where an analytic solution can be given --
as well as using the full U(1) Abelian-Higgs model \cite{dgt,agk}, where only numerical solutions
are available. In the latter case, these solutions have
been interpreted to represent black hole solutions with long range ``hair'' and are thus counterexamples
to the No hair conjecture which states that black holes are uniquely characterized by
their mass, charge and angular momentum.
Interestingly, the solution found in \cite{afv} is a Schwarzschild solution
which however differs from the standard spherically symmetric case by the replacement of the angular
variable $\phi$  by $\beta\phi$, where the parameter $\beta$ is related to the deficit
angle by $\delta=2\pi(1-\beta)$. In this sense, the space-time is thus not uniquely determined by the mass, but
is described by the mass {\it and} deficit angle parameter $\beta$. Of course, it is
then easy to extend known analytically given space--times to the conical case
by making the mentioned substitution.

In order to understand details of gravitational fields of 
massive objects and to be able to predict observational consequences, it is
important to understand how test particles move in these space-times. Next to the above mentioned
examples the complete set of solutions to the geodesic equation in the space--time of a Schwarzschild black hole pierced by a cosmic string has been given recently in \cite{hhls}.

The geodesic equation for the motion of test particles in the space--time of an uncharged, rotating black hole in 4 space--time dimensions (given by the Kerr solution) pierced by an infinitely thin string aligned with the rotation axis of the black hole has been given in \cite{gm} and
spherical orbits and the Lense--Thirring precession have been discussed.
Solutions to the geodesic equation in this space--time have also been given in \cite{Ozdemir2003}, but the test particle motion has been restricted to the equatorial plane and gravitomagnetic effects have been studied.
Moreover, small perturbations around circular orbits have been discussed in \cite{Ozdemir2004}. The motion of a test scalar quantum particle in the
space--time of a  Kerr--Newman black hole pierced by a cosmic string has been discussed in \cite{Fernandes2006} and it was observed that the presence of a cosmic string alters the 
corresponding observables.

The aim of this paper is to determine the {\it complete set} of
analytic solutions of the geodesic equations in the space--time of
a Kerr black hole pierced by a cosmic string and to derive analytical expressions for observable effects which can be used for astrophysical searches
for such cosmic strings. The accurate computation of geodesics for massive particles is important
in order to understand gravitational wave signals from binaries which can later be compared with
eventual gravitational wave measurements. The computation of geodesics for massless particles is important in order
to understand how light signals pass by black holes or other massive objects.
Moreover, our work has a direct link to the recently proposed explanation \cite{Poltis2010} of
the observed alignment of the polarization vector of quasars on cosmological scales \cite{quasars}.
In \cite{Poltis2010} the assumption is used  that two originally linked electroweak string loops decayed
via the formation of monopole--antimonopole pairs in the early universe. The remnants of this decay are interconnected
loops of magnetic field whose radii have grown due to the expansion of the universe and today should
be on the order of Gpc. Interestingly, it was found that the rotation axis of a quasar would
align with the direction of the magnetic field. Since the size of the magnetic field loops are much larger
than the size of the quasars, we can assume the loop to be approximated by a straight line of magnetic field --
that is aligned with the rotation axis of the supermassive black hole in the center of the quasar -- at the position of the quasar. If we assume our infinitely thin cosmic string to be a toy model for finite width strings (e.g.
electroweak or Abelian--Higgs strings) where the latter would have a magnetic flux along their axis, the model
studied in this paper would describe a quasar with its rotation axis equal to the axis of the magnetic
flux.

Our paper is organized as follows: in Section II, we give the geodesic equation, in Section
III, we classify the solutions, while in Section IV, we give the solutions to the geodesic
equation and in Section V we present examples of orbits. In Section VI we discuss observables
such as the perihelion shift and the Lense--Thirring effect. We conclude in Section VII.

\begin{figure}[p!]
\begin{center}
\resizebox{6in}{!}{\includegraphics{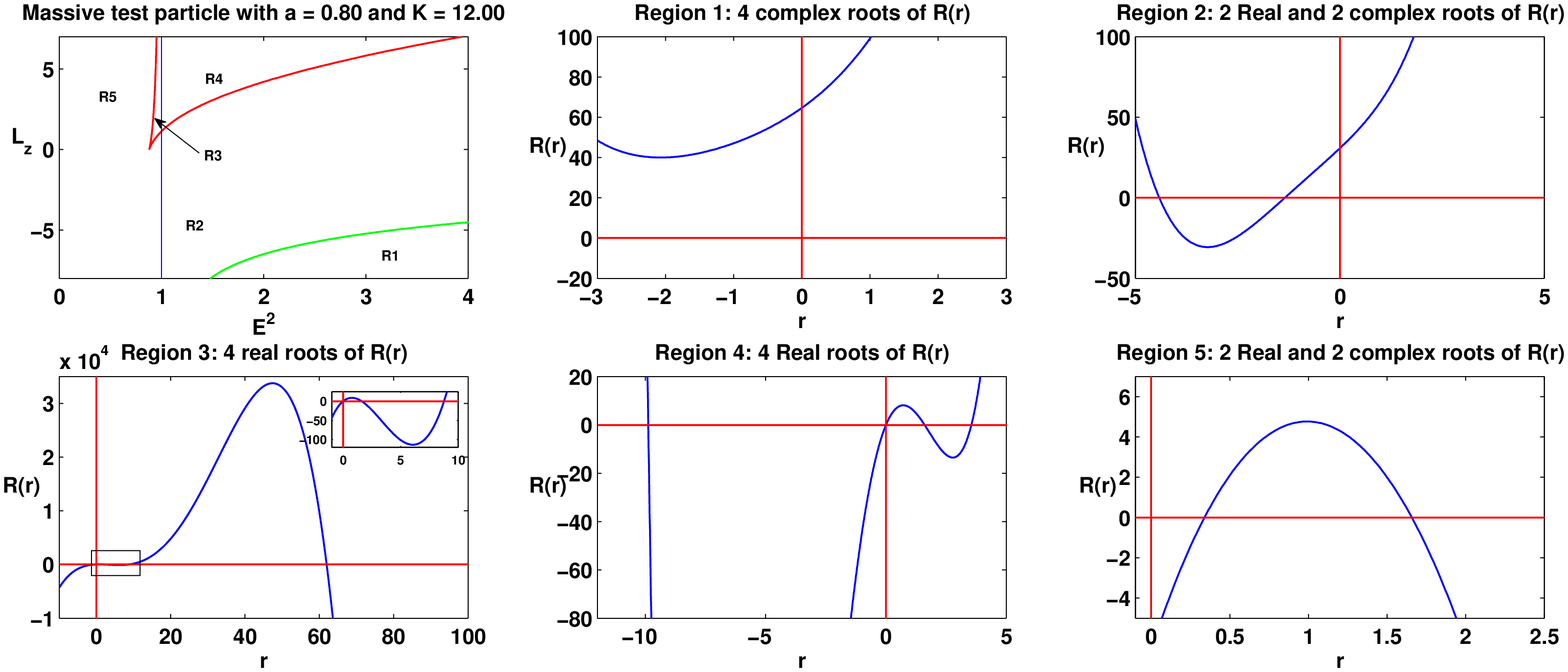}}
\end{center}
\caption{The domain of existence of solutions to the geodesic
equations for $M= 1$, $a = 0.8$ and $K=4$ that follows from the requirement $R(r) > 0$. The top left figure shows the domain of existence 
in the $L_z$-$E^2$-plane. The green and red lines represent $R(r)=0$, where the
green line is for $\kappa \in ]-\infty:0]$ and the red line for $\kappa \in [0:+\infty[$.
The remaining figures show the function $R(r)$ for
the different domains denoted by $R1$ to $R5$. }\label{LzER}
\end{figure}

\begin{figure}[p!]
\centering
\subfigure[varying Carter constant $K$ for $M=1$, $a= 0.8$.]{
\includegraphics[scale=0.4]{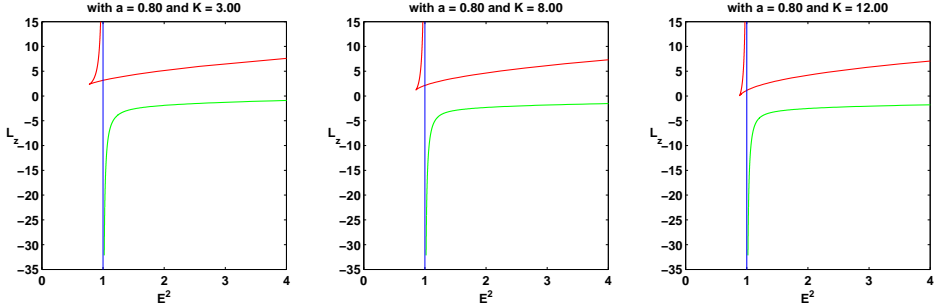}
\label{LzERvK}
}

\subfigure[varying $a$ for $M=1$, $K= 3.0$.]{
\includegraphics[scale=0.4]{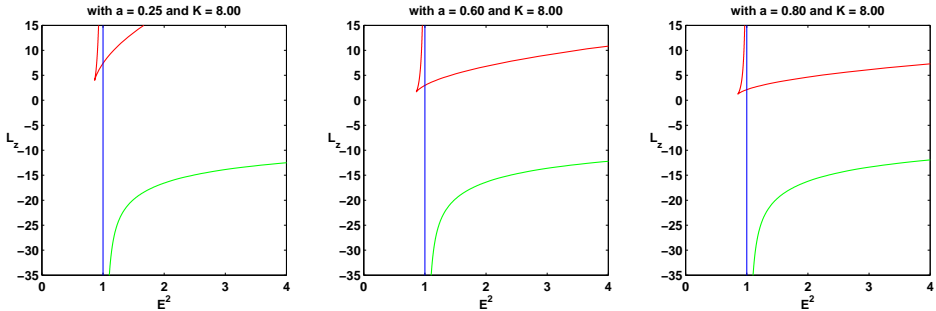}
\label{LzERva}
}
\caption[Optional caption for list of figures]{The change of the $L_z$-$E^2$ plot for varying Carter constant \subref{LzERvK} and angular momentum per unit mass of the black hole $a$ \subref{LzERva}, respectively.}\label{LzERv}
\end{figure}

\section{The geodesic equation}
We consider the geodesic equation
\begin{equation}
 \frac{d^2 x^{\mu}}{d\tau^2} + \Gamma^{\mu}_{\rho\sigma} \frac{dx^{\rho}}{d\tau}\frac{dx^{\sigma}}{d\tau}=0  \ ,
\end{equation}
where $\Gamma^{\mu}_{\rho\sigma}$ denotes the Christoffel symbol given by
\begin{equation}
 \Gamma^{\mu}_{\rho\sigma}=\frac{1}{2}g^{\mu\nu}\left(\partial_{\rho} g_{\sigma\nu}+\partial_{\sigma} g_{\rho\nu}-\partial_{\nu} g_{\rho\sigma}\right)
\end{equation}
and $\tau$ is an affine parameter such that for time--like geodesics $d\tau^2=g_{\mu\nu}dx^{\mu} dx^{\nu}$ corresponds to proper time.

The explicit form of the metric that we are studying in this paper reads
\begin{eqnarray}\label{KCSmetric}
ds^2&=&-\left(1-\frac{2Mr}{\rho^2}\right)dt^2+\frac{\rho^2}{\Delta}dr^2+\rho^2 d\theta^2+\beta^2\left( r^2+a^2+\frac{2Mra^2\sin^2{\theta}}{\rho^2}\right)\sin^2{\theta}d\phi^2-\beta\frac{4 Mra\sin^2\theta}{\rho^2}dtd\phi \ ,
\end{eqnarray}
where $\rho^2=r^2+a^2\cos^2{\theta}$, $\Delta=r^2-2Mr+a^2$, $a=J/M$  is the angular momentum $J$ per mass $M$ of the black hole and  $0 < \beta < 1$ is the deficit angle parameter that is related to the deficit
angle $\delta=2\pi(1-\beta)$.  This metric
describes a Kerr black hole pierced by an infinitely thin cosmic string
that is aligned with the rotation axis of the black hole. The deficit angle appears due to the presence
of the cosmic string and can be expressed in terms of the 
energy per unit length $\mu$ of the cosmic string~: $\delta=8\pi G\mu\sim 8\pi (\eta/M_{\rm Pl})^2$, where
$\eta$ is the typical symmetry breaking scale at which the cosmic string formed and $M_{\rm Pl}$ is the Planck mass.
Note that we are using units such that $G=c=1$.

For $a=0$ this metric reduces to the Schwarzschild solution pierced by an infinitely thin cosmic string \cite{hhls}, while for $\beta=1$ we recover the standard Kerr solution.

Surfaces with $\Delta=0$ correspond to horizons of the Kerr solution with horizon radius
$r_{\pm}=M\pm\sqrt{M^2-a^2}$, where the $+$ sign corresponds to the event horizon, while the
$-$ sign corresponds to the Cauchy horizon.
Surfaces with $2Mr=\rho^2$ are the static limit and define a Killing horizon. The domain between the event horizon and the
static limit is the ergosphere.
In the following, we are only interested in non--extremal black hole solutions, i.e. in solutions
with $M^2 > a^2$. Note that the localization of the horizons
and the static limit, respectively are not altered by the presence of the cosmic string.

The Boyer-Lindquist coordinates ($r$, $\theta$, $\phi$) are related to cartesian coordinate ($x$, $y$, $z$) by
$x=\sqrt{r^2+a^2}\sin{\theta}\cos{(\beta\phi)}$,
$y=\sqrt{r^2+a^2}\sin{\theta}\sin{(\beta\phi)}$ and
$z=r\cos{\theta}$.
Hence $r=0$ corresponds to a disc with a deficit angle $\delta$, while
the physical singularity at $r=0$, $\theta=\pi/2$ is a ring with
a deficit angle. It is then clear -- in contrast to the $a=0$ limit -- that negative values of $r$ are allowed. When crossing $r=0$ to negative
values of $r$ one enters into another conical space--time that however possesses no horizons. The Penrose diagram of this
space--time looks exactly like that of the standard Kerr space--time \cite{ONeill}, however each point in the diagram would correspond to a sphere
with deficit angle $\delta$.

The Lagrangian reads
\begin{eqnarray}
\mathcal{L}&=&\frac{1}{2}g_{\mu\nu}\frac{dx^{\mu}}{d\tau}\frac{dx^{\nu}}{d\tau}=\frac{1}{2}\varepsilon  \ , \end{eqnarray}
where  $\varepsilon=-1$ for massive test particles and $\varepsilon=0$ for 
massless test particles, respectively. The Killing vectors of this space--time are $\frac{\partial}{\partial t}$
and $\frac{\partial}{\partial \phi}$. The constants of motion are the energy $E$ and the angular momentum $L_{z}$ which are given by the generalized momenta $p_{t}$ and $p_{\phi}$
\begin{eqnarray}
-p_{t}&=&-\frac{\partial\mathcal{L}}{\partial\dot{t}}=-\dot{t}g_{tt} -\dot{\phi}g_{t\phi}=\left(1-\frac{2Mr}{\rho^2}\right)\dot{t} + 
\beta\frac{2Mar}{\rho^2} \sin^2\theta\dot{\phi}  =:E  \ \ \ , \\
p_{\phi}&=&\frac{\partial\mathcal{L}}{\partial\dot{\phi}}=\dot{t}g_{t\phi} +\dot{\phi}g_{\phi\phi}=-\beta\frac{2Mar}{\rho^2}\sin^2\theta \dot{t} + 
\beta^2\frac{(r^2+a^2)^2 - \Delta a^2 \sin^2\theta}{\rho^2}\sin^2\theta \dot{\phi}  =:\beta L_z  \ .
\end{eqnarray}
The dot denotes the differentiation with respect to the affine parameter $\tau$. 
Therefore the presence of a cosmic string aligned with the rotation axis of the Kerr black hole decreases the magnitude of the generalized momenta due to the parameter $\beta$ 
as compared to the standard Kerr space--time. In particular, for a given energy
$E$ a particle on the surface defined by $g_{tt}=0$ (the static limit) has larger $\dot{\phi}$ as compared to the
standard Kerr case.

There is another constant of motion, namely the Carter constant $K$ \cite{Carter1968} that appears when separating the Hamilton--Jacobi equations
\begin{eqnarray}\label{HJ}
\frac{\partial S}{\partial\tau}&=&\frac{1}{2}g^{\mu\nu}\left(\partial_{\mu}S\right)\left(\partial_{\nu}S\right) \ ,
\end{eqnarray}
 which persists in the presence
of a cosmic string. $S$ denotes the principal function of Hamilton for which we make the following
Ansatz
\begin{eqnarray}\label{Sfunc}
S=\frac{1}{2}\varepsilon\tau-Et+\beta L_{z}\phi+S_{r}(r)+S_{\theta}(\theta) \ ,
\end{eqnarray}
where $S_r(r)$ and $S_{\theta}(\theta)$ are functions of $r$ and $\theta$ only, respectively.

By differentiating $S$ with respect to the constants $\varepsilon$, $E$, $L_z$ and $\mathcal{Q}:=K-(L_z-aE)^2$, where
${\cal Q}$ is the modified Carter constant,
we find the geodesic equations
\begin{eqnarray}
\rho^2 \dot{r}&=&\pm\sqrt{R(r)}\label{rdot} \ , \\
\rho^2 \dot{\theta}&=&\pm\sqrt{\Theta(\theta)}\label{thetadot} \ ,\\
\rho^2\beta \dot{\phi}&=&(L_z\csc^{2}{\theta}-aE)+\frac{aP(r)}{\Delta(r)}\label{phidot} \ ,\\
\rho^2 \dot{t}&=&a(L_z-aE\sin^2{\theta})+\frac{(r^2+a^2)P(r)}{\Delta(r)}\label{tdot}  
\end{eqnarray}
with
\begin{eqnarray}
\Theta(\theta)&=&\mathcal{Q}-\cos^2{\theta}\left(L_{z}^{2}\csc^{2}{\theta}-a^2(E^{2}+\varepsilon)\right)\label{Theta}
 \ \ , \ \ \\
P(r)&=&E(r^2+a^2)-L_{z}a\label{Pr}\ \ , \ \ \\
R(r)&=&P(r)^{2}-\Delta(r)\left(\mathcal{Q}+(L_{z}-aE)^{2}-\varepsilon r^{2}\label{Rr}\right) \ .
\end{eqnarray}
Introducing a new parameter, the so-called {\it Mino time} \cite{Mino2003} given by
$d\lambda=\frac{d\tau}{\rho^2}$
the $r$- and $\theta$-component of the geodesic equation can be decoupled
\begin{eqnarray}
\frac{dr}{d\lambda}&=&\pm\sqrt{R(r)}\label{rlambda} \ , \ \ \\
\frac{d\theta}{d\lambda}&=&\pm\sqrt{\Theta(\theta)} \label{thetalambda} \ , \\ \frac{d\phi}{d\lambda}&=&\frac{1}{\beta}\left(\frac{L_z\csc^{2}{\theta}-aE}{\sqrt{\Theta(\theta)}} \frac{d\theta}{d\lambda}+\frac{aP(r)}{\Delta(r)\sqrt{R(r)}} \frac{dr}{d\lambda}\right)\label{philambda} \ , \\
\frac{dt}{d\lambda}&=&\frac{a(L_z-aE\sin^2{\theta})}{\sqrt{\Theta(\theta)}}\frac{d\theta}{d\lambda}+\frac{(r^2+a^2)P(r)}{\Delta(r)\sqrt{R(r)}}\frac{dr}{d\lambda}\label{tlambda}  \ .
\end{eqnarray}
For $a=0$ these equations reduce to the equations of motion in a space--time of a
Schwarzschild black hole pierced by a cosmic string \cite{hhls}, while for 
$\beta=1$ we recover the geodesic equation in the Kerr space--time.
As initial conditions we will choose the $+$-signs in (\ref{rlambda}) and (\ref{thetalambda}).

\section{Classification of solutions}
The classification of the solutions of the geodesic equations (\ref{rlambda})-(\ref{tlambda}) can be done
with respect to the modified Carter constant $\mathcal{Q}$, the mass of the black hole $M$, the
angular momentum per unit mass of the black hole $a$ as well
as the energy $E$ and angular momentum $L_z$ of the massive ($\varepsilon=-1$) or
massless $(\varepsilon=0$) test particle.
Apparently, the deficit parameter $\beta$ does not appear in the $r$, $\theta$ and $t$ component
of the geodesic equation and will hence only influence the $\phi$-motion.

\subsection{$r$--motion}
In order to have solutions of the geodesic equation for $r$ we have to require $R(r) > 0$.
Therefore solutions of the $r$--component of the geodesic equation
exist only for specific choices of $E$ and $L_z$.  $R(r)$ will have either 4 real, 2 real and 2 complex
or 4 complex zeros. On the boundaries between the domains in the $E^2$--$L_z$--plane corresponding
to these three different possibilities $R(r)$ necessarily has a double zero. In order to find
these boundaries we make the Ansatz $R(r)=(r-\kappa)^2 \left((E^2+\varepsilon)r^2 + \rho_1 r + \rho_2\right)$ where
$\kappa$, $\rho_1$ and $\rho_2$ have to be determined. We find that for $R(r)=0$ we have
the following parametric expressions for $E$ and $L_z$
\begin{eqnarray}
E(\kappa)&=&\frac{-2\kappa^3-3\varepsilon M \kappa^2+(\varepsilon a^2)\kappa-KM}{\sqrt{P_\kappa}} \ ,\\
L_z(\kappa)&=&\frac{\varepsilon M \kappa^4+(K-2a^2-\varepsilon a^2)\kappa^3-(3MK+3\varepsilon Ma^2)\kappa^2+(\varepsilon a^4+a^2K)\kappa+a^2MK}{a\sqrt{P_\kappa}} \ ,
\end{eqnarray}
where
\begin{eqnarray}
P_\kappa&=&4\kappa^6+8\varepsilon M \kappa^5+(4K-4a\varepsilon)\kappa^4-8MK\kappa^3+4a^2K\kappa^2  \ .
\end{eqnarray}
Note that $r=\kappa$ corresponds to a double zero of $R(r)$ and hence represents spherical
orbits. 

Examples for the polynomial $R(r)$ are given in Fig.\ref{LzER} and Fig \ref{LzERv}.
In Fig.\ref{LzER} we show the different domains in the $E^2$-$L_z$-plane for 
$M=1$, $a=0.8$ and $K=4$. The blue line corresponds to $E^2=1$, the green line
to $\kappa \in ]-\infty,0]$ and the red line to $\kappa \in [0,+\infty[$. The plots of $R(r)$ in
the domains R1 to R5 are also shown. For R1 $R(r)$ possesses 4 complex zeros,
for $R2$ and $R5$ 2 real and 2 complex zeros, while for R3 and R4 there are four real zeros.
The following type of orbits are then possible
\cite{ONeill}~:
\begin{itemize}
\item \textbf{Flyby orbit:} the test particle starts from $\pm\infty$, reaches a minimal value of  $|r|$ and flies back to  $r$ = $\pm\infty$. There are two flyby orbits in R2 and R4, respectively.
\item\textbf{Transit orbit:} the test particle starts from $\pm\infty$, crosses $r$ = 0 and continues to  $r$ = $\mp\infty$. There is a transit orbit in R1.
\item \textbf{Bound orbit:} the test particle oscillates in an interval [$r_2,r_1$] where $r_1$ $>$ $r_2$. There are two bound orbits in R3 and one bound orbit in R4 and R5, respectively.
\item \textbf{Spherical orbit:} this is a special bound orbit with $r$ constant. $r=\kappa$ is a spherical orbit that can be stable or unstable. 

\end{itemize}

In Fig.\ref{LzERv} we show how the domains R1 to R5 change when changing the Carter constant $K$ (upper three figures).
Obviously R3 decreases in size when increasing $K$. We also give the change of R1 to R5 for changing
angular momentum per unit mass $a$ (lower three figures). In this case, the domain R4 is increasing for increasing $a$.

\subsection{$\theta$--motion}
\begin{figure}[p!]
\centering
\subfigure[ $M$ = 1, $a$ = 0.8 with the change of $K$]{
\includegraphics[scale=0.4]{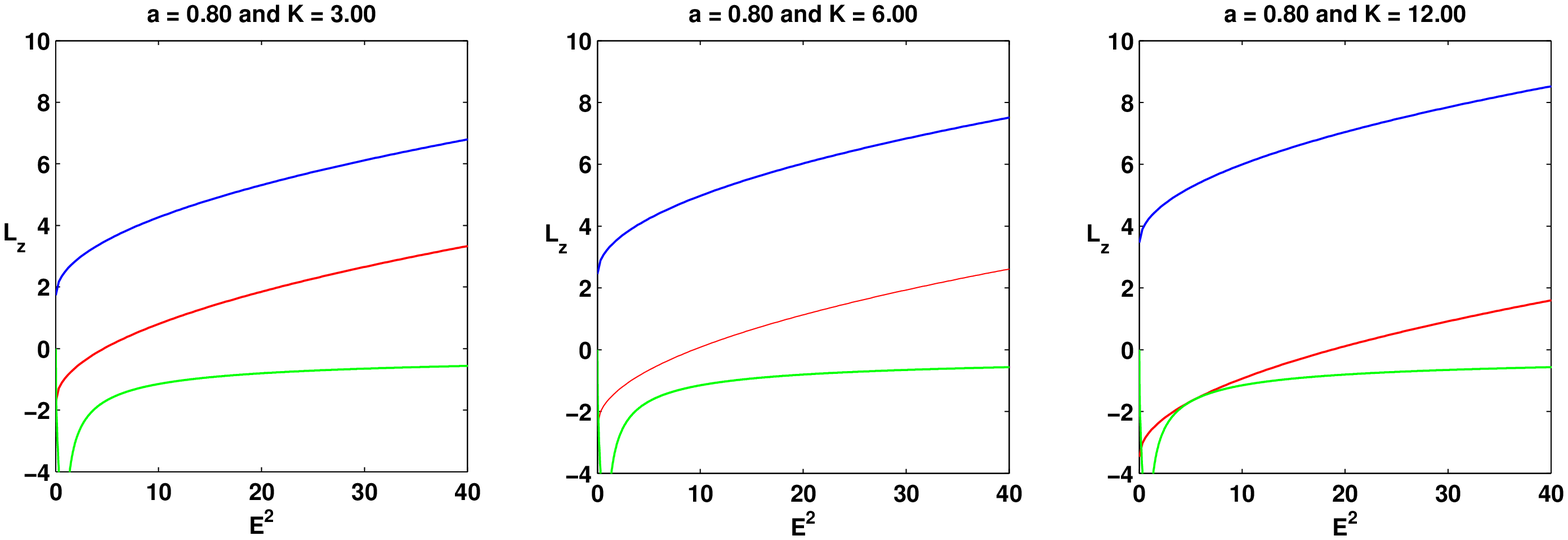}
\label{LzETvK}
}
\subfigure[ $a$ at $M$ = 1, $K$ = 3.0 with the change of $a$]{
\includegraphics[scale=0.4]{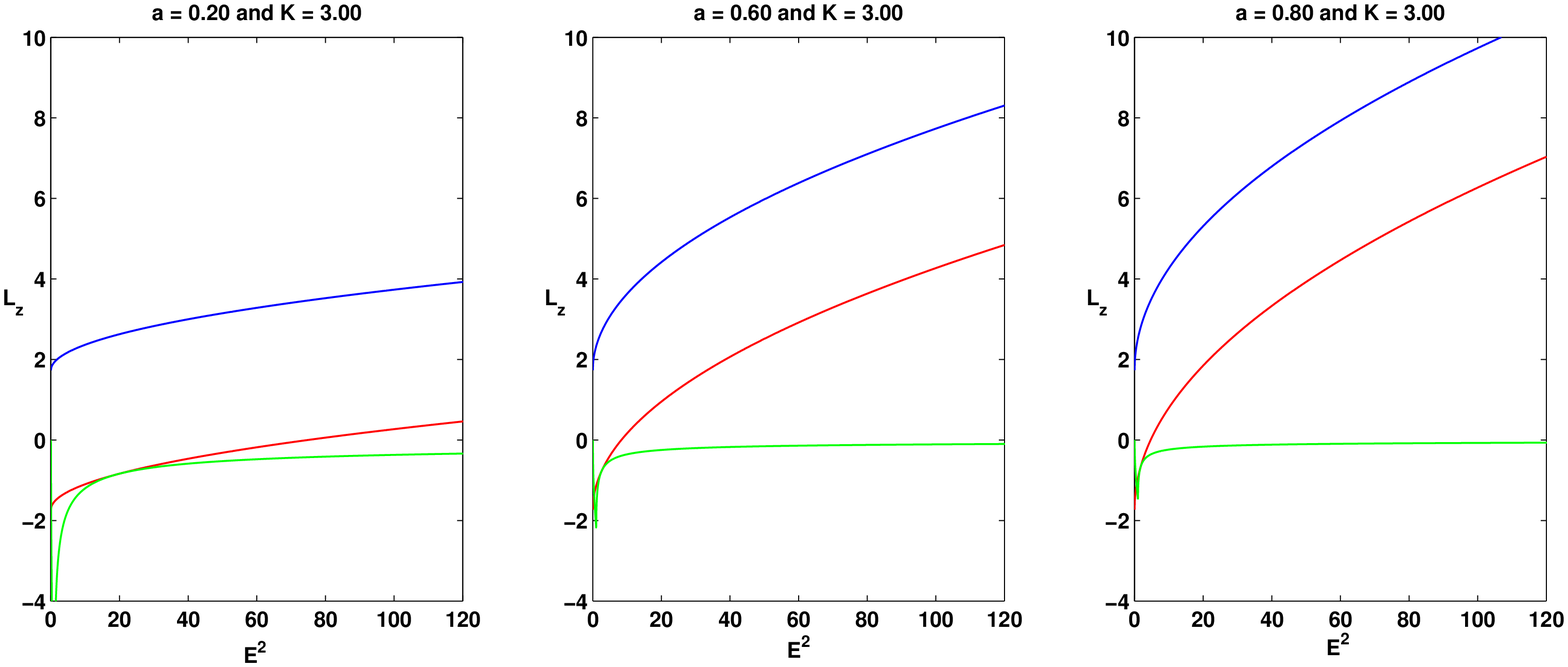}
\label{LzETva}
}

\caption[Optional caption for list of figures]{The change of the domain of existence of solutions
to the geodesic equations
in the $L_z$-$E^2$ plane for several values of the Carter constant $K$ and $M=1$, $a=0.8$ \subref{LzETvK} and several values of the angular momentum per unit mass of the black hole $a$
and $M=1$ and $K=3$ \subref{LzETva}, respectively. Note that solutions
exist only below the blue, above the red and above the green line.}\label{LzETv}
\end{figure}

It is obvious from the form of the geodesic equations that we should have
$\Theta(\theta) > 0$, i.e. $\theta$--motion is allowed only for those
$\theta$ for which $\Theta(\theta) > 0$.
This in turn means that when we fix $\mathcal{Q}$ and $a$ only particular
values of $E$ and $L_z$ are allowed. 
\begin{enumerate}
 \item equatorial motion with $\theta=\pi/2$ for massive and massless particles: in this case, it follows
from (\ref{Theta}) that $\mathcal{Q}= 0$, i.e.
\begin{equation}
 L_z = aE\pm \sqrt{K}\ .
\end{equation}
This is shown in Fig.\ref{LzETvK} and Fig.\ref{LzETva} where the blue and the red line 
indicate $L_z=aE+\sqrt{K}$ and $L_z=aE-\sqrt{K}$, respectively. 
\item polar motion with $\theta=0$ or $\theta=\pi$: obviously in this case, we have to choose
$L_z=0$ and $\mathcal{Q} \ge -a^2(E^2+\varepsilon)$.
\item motion for $0 < \theta < \pi$ with $\theta\neq \pi/2$: in this case, we find that (see Appendix for details)
 \begin{equation}
\label{lzmassive}
L_z \ge \frac{1}{2}\frac{\left(E-\sqrt{E^2-\varepsilon}\right)\left(a^2\varepsilon + K\right)}{\varepsilon a}
\end{equation}
for massive particles and
\begin{equation}
\label{lzmassless}
L_z \ge \frac{K}{4aE^2}
\end{equation}
for massless particles.
This is shown in Fig.\ref{LzETvK} and Fig.\ref{LzETva} where the green line indicates
$L_z = \frac{1}{2}\left(E-\sqrt{E^2-\varepsilon}\right)\left(a^2\varepsilon + K\right)/(\varepsilon a)$.
The allowed domain in the $L_z$-$E^2$-plane is the one above this green line.
\end{enumerate}


\begin{figure}[p!]
\begin{center}
\resizebox{6in}{!}{\includegraphics{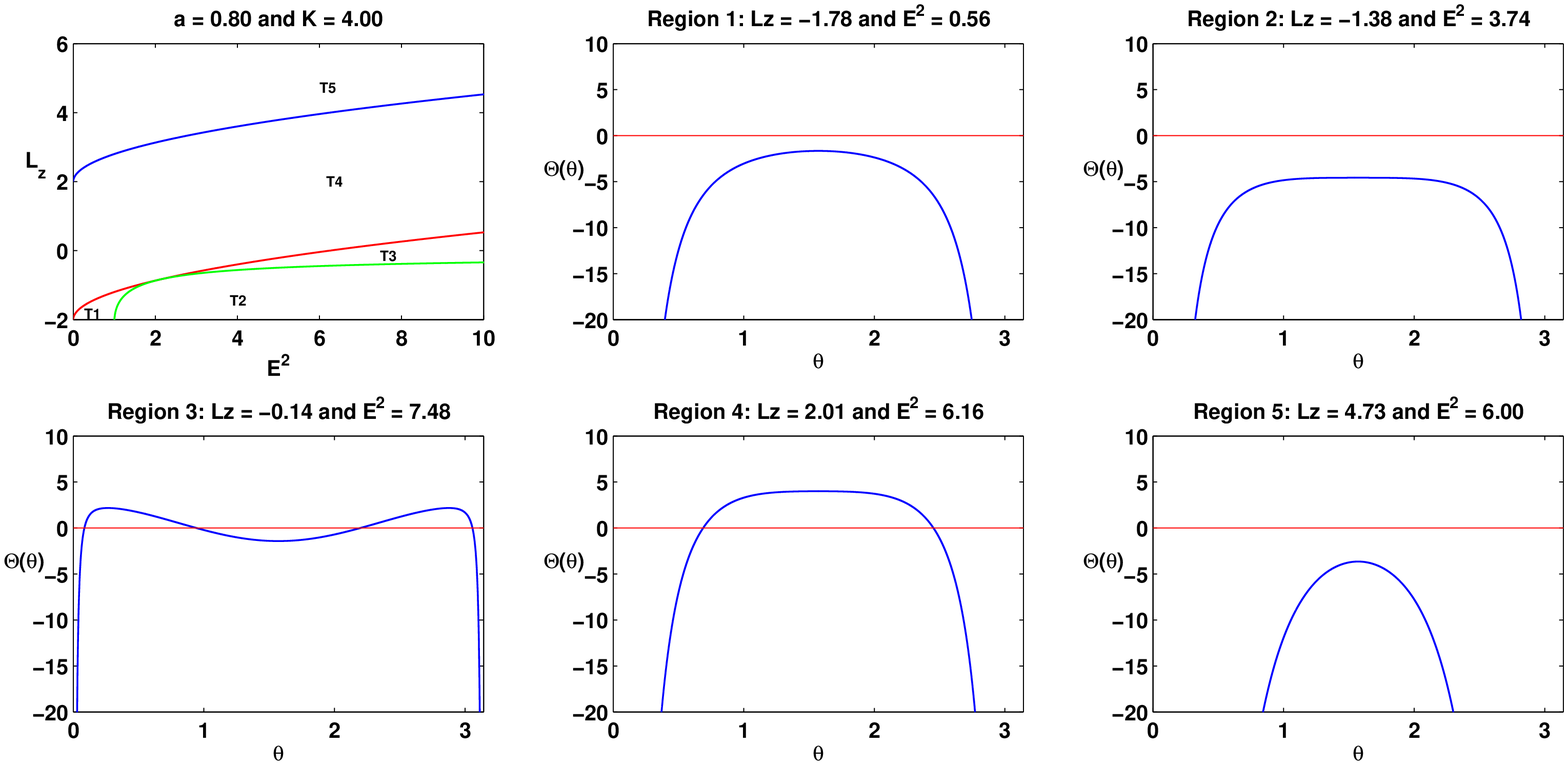}}
\end{center}
\caption{The domain of existence of solutions to the geodesic
equations for $M= 1$, $a = 0.8$ and $K=4$ that follows from the requirement $\Theta(\theta) > 0$. The top left figure shows the domain of existence 
in the $L_z$-$E^2$-plane. The remaining figures show the function $\Theta(\theta)$ for
the different domains denoted by $T1$ to $T5$. Solutions exist only for parameters $L_{z}$ and $E$  in domains $T3$ and $T4$, respectively.}\label{LzET}
\end{figure}

We show an example of the different domains corresponding to zeros of $\Theta(\theta)$ in Fig.\ref{LzET} for $M=1$, $a=0.8$ and $K=4$.
The function $\Theta(\theta)$ is plotted for the domains denoted $T1$ to $T5$. In domains $T1$, $T2$ and $T5$
the polynomial $\Theta(\theta)$ has no real zeros, while it has four real zeros in domain $T3$ and two real
zeros in domain $T4$, respectively.
Clearly, only
in the domains $T3$ and $T4$ solutions to the geodesic equations exist.
In $T3$ there are two domains in $\theta$ for which $\Theta(\theta) >0$. Here the particles oscillate between $\theta_{\rm min}$ and $\theta_{\rm max}$ but cannot cross $\theta$ = $\frac{\pi}{2}$. In domain $T4$ the particle can cross the equatorial plane $\theta=\pi/2$.

\section{Analytical solutions of the geodesic equation}
In the following, we will present the analytical solutions of the
geodesic equation for the  $r$, $\theta$, $\phi$ and $t$ component.
We will present these solutions using Mino time $\lambda$.

\subsection{$r$--motion}\label{4r2r2c}
The polynomial $R(r)$ can be written in the form
\begin{eqnarray}
R(r)&=&\left( {E}^{2}+\varepsilon \right)\prod_{j=1}^{4}(r-r_{j})  \ ,
\end{eqnarray}
where the $r_{j}$ denote the zeros of $R(r)$ with $r_{1}$ the largest real zero.
Using the coordinate
transformation $r=1/y+r_1$ the polynomial $R(r)$ can be transformed to a 3rd order polynomial in $y$ using
\begin{equation}
d\lambda= \frac{dr}{\sqrt{R(r)}}=-\frac{dy}{\sqrt{\left(E^2+\varepsilon\right)\sum_{i=0}^3 b_i y^i}}  \ 
\end{equation}
with 
\begin{eqnarray}
b_{0}&=&1  \ ,\\
b_{1}&=&\left( -{ r_2}-{ r_4}+3{ r_1}-{ r_3} \right) \ ,\\
b_{2}&=&\left( { r_2}{ r_4}-2{ r_1}{r_2}+{ r_2}{ r_3}+3{{ r_1}}^{2}-2{ r_1}{ r_3}-2{ r_1}{ r_4}+{ r_3}{ r_4} \right) \ ,\\
b_{3}&=&\left( -{ r_3}{{ r_1}}^{2}+{ r_1}{ r_2}{ r_3}+{r_1}{ r_2}{ r_4}+{ r_1}{ r_3}{ r_4}-{ r_2}{{r_1}}^{2}+{{ r_1}}^{3}-{ r_4}{{ r_1}}^{2}-{r_2}{r_3}{ r_4} \right)  \ .
\end{eqnarray}

Introducing the variable
\begin{equation}
\label{zgammaalpha}
z=\frac{y-\alpha}{\gamma} \ \ \ {\rm with} \ \ \  \ \
\gamma =\sqrt[3]{\frac{4}{b_{3}({E}^{2}+\varepsilon )}} \ \ , \ \  \alpha =-\frac{b_{2}}{3b_{3}} 
\end{equation}
we can write the solution of (\ref{rlambda}) as follows~:
\begin{eqnarray}
r(\lambda)&=&\left(\frac{1}{\gamma\wp\left(\frac{1}{\gamma}(\lambda-\lambda_{0})+C_{r};\tilde{g}_2,\tilde{g}_3\right)+\alpha}+r_{1}\right)  \ .
\end{eqnarray}
The integration constant is given by
\begin{eqnarray}
\label{cr}
C_{r} = \int_{z_0}^{\infty}\frac{dz}{\sqrt{4z^{3}-\tilde{g}_2z-\tilde{g}_3}}\label{zsol}  \ ,
\end{eqnarray}
where $z_0$ denotes the value of $z$ that corresponds to the initial radius and
 \begin{equation}
\tilde{g}_{2}=-{\frac {{2}^{2/3}\sqrt [3]{(E^{2}+\varepsilon)^{2}{{ b_3}}^{2}} \left( -{{
b_2}}^{2}+3{ b_1}{ b_3} \right) }{3{{ b_3}}^{2}}} \ \ \ , \ \ \  \tilde{g}_{3}=-{\frac {(E^{2}+\varepsilon) \left( 2{{ b_2}}^{3}+27{ b_0}{{ b_3}}^{2}-9{ b_1}{ b_2}{ b_3} \right) }{27{{ b_3}}^{2}}} \ .
\end{equation}

Depending on the sign of the discriminant $\tilde{D}=\tilde{g}_2^3 - 27 \tilde{g}_3^2$ we have different values for $C_r$. In the following, we will denote the zeros
of the 3rd order polynomial in the square zero of (\ref{cr}) as $\tilde{e}_1 > \tilde{e}_2 > \tilde{e}_3$.

\begin{enumerate}
 \item $\tilde{D} > 0$~: In this case $R(r)$ has four real zeros which corresponds to domains
$R3$ and $R4$ in Fig.\ref{LzER}.  

For the outer bound orbit in domain R3 with minimal radius $r_2$ and
maximal radius $r_1$  we choose
$z_0=\tilde{e}_1$ such that $C_r=K(\mathcal{K})/\sqrt{\tilde{e}_1-\tilde{e}_3}=:\tilde{\omega}_1$
where $K(\mathcal{K})$ is the complete elliptic integral of the first kind and $\mathcal{K}=\sqrt{{\tilde{e}_2-\tilde{e}_3}}/\sqrt{{\tilde{e}_1-\tilde{e}_3}}$ is the modulus of the elliptic integral.
For the inner bound orbit in domain R3 with minimal radius $r_4$ and
maximal radius $r_3$ we choose
$z_0=\tilde{e}_2$ such that $C_r=K(\mathcal{K})/\sqrt{\tilde{e}_1-\tilde{e}_3}+
iK(\mathcal{K}')/\sqrt{\tilde{e}_1-\tilde{e}_3}=:\tilde{\omega}_1+\tilde{\omega}_2$
where $\mathcal{K}'=\sqrt{1-\mathcal{K}^2}$.
For the flyby orbit in R4 with initial minimal radius $r_1$, we choose
$z_0=\infty$ such that $C_r=0$, while for the inner bound orbit in R4 with
initial maximal radius $r_2$ and minimal radius $r_3$ we choose $z_0=\tilde{e}_2$ such that
$C_r=\tilde{\omega}_1+\tilde{\omega}_2$.

\item $\tilde{D} < 0$~: In this case $R(r)$ has two real zeros which corresponds
to domains R2 and R5 in Fig.\ref{LzER}. In the following, we will denote the real zero
of the 3rd order polynomial in the square zero of (\ref{cr}) as $\tilde{e}_2$, while
the two complex zeros are denoted by $\tilde{e}_1$ and $\tilde{e}_3$.

For the flyby orbit in domain R2 with initial minimal radius $r_1$ we choose $z_0=\infty$ such
that $C_r=0$. For the bound orbit in domain R5 with initial maximal radius $r_1$ we also
choose $z_0=\infty$, i.e. $C_r=0$.

\item $\tilde{D} = 0$~: In this case $R(r)$ has one real zero. The orbit
in this case will be a bound spherical orbit.
\end{enumerate}

Note that there is also the possibility of four complex zeros.
This however would simply correspond to a transit orbit
with $r=-\infty$ $\rightarrow +\infty$. We will not discuss
this case in detail in this paper.

\subsection{$\theta$--motion}\label{thetasection}
The solution of (\ref{thetalambda}) as a function of Mino time $\lambda$ is given by
\begin{eqnarray}
\theta(\lambda)&=&\arccos\left[\pm\frac{1}{\sqrt{\mu\wp\left(\frac{1}{\mu}(\lambda-\lambda_{0})+C_{\theta};g_2,g_3\right)+\nu}}\right]\label{thetasol}    \ \  ,
\end{eqnarray}
where 
\begin{equation}
\label{munu}
\mu={\mathcal Q}^{-1/3}  \ \ \ ,  \ \ \ \nu=\left(\mathcal{Q}+L_z^2 - a^2(E^2+\varepsilon)\right)/(3\mathcal{Q}) \ .
\end{equation}
The positive (negative) sign of the $\arccos$ corresponds to the choice $\theta > \pi/2$ ($\theta < \pi/2$). For the $\theta$ motion
in domain $T4$, the solutions thus have to be ``glued together'' at $\theta=\pi/2$.

The integration constant $C_{\theta}$ is given by 
\begin{equation}
C_{\theta}=\int\limits_{w_0}^{\infty} \frac{dw}{\sqrt{4w^3-g_2w-g_3}}  \ ,
\end{equation}
where $w$ is related to $\theta$ by $\cos^2\theta=(\mu w + \nu)^{-1}$, $w_0$ corresponds
to the value of $w$ that represents the initial $\theta$ and 
\begin{equation} g_{2}=-\frac{2^{2/3}(a_{1}^2)^{1/3}(3a_{1}a_{3}-a_{2}^2)}{3a_{1}^{2}} \ \ \ , \ \ \  g_{3}=-\frac{2a_{2}^{3}-9a_1a_2a_3}{27a_1^2} \ ,
 \end{equation}
where $a_1=4\mathcal{Q}$, $a_2=4\left(a^2(E^2+\varepsilon)-L_z - \mathcal{Q}\right)$ and
$a_3=-4a^2(E^2+\varepsilon)$.
The discriminant $D=g_2^3-27g_3^2$ is always positive, so we have three real zeros of the
3rd order polynomial, which we call $e_1 > e_2 > e_3$ in the following. Note that 
though we might have three real zeros for the polynomial in $w$ these zeros might
not fulfill $(\mu w + \nu)^{-1}=\cos^2\theta \le 1$. Moreover to each zero of the polynomial
in $w$ correspond two values of $\theta$ fulfilling $\cos\theta=\pm (\mu w + \nu)^{-1/2}$.
Hence, depending on the values of $e_1$, $e_2$ and $e_3$ the polynomial $\Theta(\theta)$ can have
four, two or no real zeros.

In T4, we typically choose $w_0=e_1$ such that $C_{\theta}=K(\mathcal{K})/\sqrt{e_{1}-e_3}=:\omega_{1}$, where
$K(\mathcal{K})$ is the complete elliptic integral of the first kind and $\mathcal{K}=\sqrt{{e_2-e_3}}/\sqrt{{e_1-e_3}}$ is the modulus of the elliptic integral.

In T3, we typically choose $w_0=e_3$ such that $C_{\theta}=iK(\mathcal{K}')/\sqrt{e_{1}-e_3}=:\omega_{2}$, where
$K(\mathcal{K}')=\sqrt{1-\mathcal{K}^2}$.

\subsection{$\phi$--motion}\label{phimotion}
The equation for the $\phi$ component (\ref{philambda}) is separated into a $\theta$-dependent and $r$-dependent part
\begin{eqnarray}
\beta d\phi=\frac{L_z\csc^{2}{\theta}-aE}{\sqrt{\Theta(\theta)}}d\theta +\frac{a\Delta^{-1}P(r)}{\sqrt{R(r)}}d r=:dI_{\theta}+dI_{r}\label{phiItIr}   \ .
\end{eqnarray}
The solutions for $I_{\theta}$ and $I_r$ are (see Appendix for more details)~:
\begin{eqnarray}
 I_{\theta}&=&
(L_{z}-aE)(\lambda-\lambda_{0})+\sum_{i=1}^{2}\frac{\nu+\mu d_{0}}{\wp'(x_{i})}\left[\frac{(\lambda-\lambda_{0})}{\mu}\zeta(x_{i})+\ln{(\sigma(x-x_i))}-\ln{(\sigma(x_0-x_i))}\right]\ , \\
I_{r}&=&-K_{0}(\lambda-\lambda_{0})+\sum_{i,j=1}^2\frac{ K_{j}}{\wp'(u_{ji})}\left[\frac{-(\lambda-\lambda_{0})}{\gamma}\zeta(u_{ji})+\ln{(\sigma(u-u_{ji}))}-\ln{(\sigma(u_{0}-u_{ji}))}\right] \ ,
\end{eqnarray}
where $\zeta$ and $\sigma$ denote the Weierstrass zeta-- and sigma--function, respectively.
$\mu$ and $\nu$ are the variables defined in (\ref{munu}), $d_0=(1-\nu)/\mu$ and
$x=\lambda/\mu$. Moreover, we have introduced the variable
$u=-\left(\frac{1}{\gamma}(\lambda-\lambda_{0})+C_r\right)$ with $u(\lambda_{0})=-C_r$ and
the $\gamma$ given in (\ref{zgammaalpha}). In addition $u_{ji}$ =$e_j$ with $\wp(u_{11})$ = $\wp(u_{12})$ = $e_1$ and $\wp(u_{21})$ = $\wp(u_{22})=e_2$. Finally, the $K_j$ appear when rewriting
$dI_r$ as follows
\begin{eqnarray}
dI_{r}&=&K_0\frac{\gamma dz}{\sqrt{4z^{3}-\tilde{g}_2z-\tilde{g}_3}}+\sum_{j=1}^{2}K_{j}\frac{dz}{(z-e_{j})\sqrt{4z^{3}-\tilde{g}_2z-\tilde{g}_3}}\ \ \ , 
\end{eqnarray}
where $e_{j}:=\frac{y_j-\alpha}{\gamma}$ and $y=(r-r_1)^{-1}$.

\subsection{$t$--motion}\label{tmotion}
From (\ref{tlambda}) we have
\begin{eqnarray}
dt=a(L_z-aE\sin^2{\theta})\frac{d\theta}{\sqrt{\Theta(\theta)}}+(r^2+a^2)\Delta(r)^{-1}P(r)\frac{dr}{\sqrt{R(r)}}
=:d\bar{I}_{\theta}+d\bar{I}_{r}.\label{dtdI}
\end{eqnarray}
To find analytical expressions for $\bar{I}_{\theta}$ and $\bar{I}_r$
we proceed as in the case of $I_{\theta}$ and $I_r$, respectively (see also Appendix) and find
\begin{eqnarray}
\bar{I}_{\theta}=a(L_z-aE)(\lambda-\lambda_0)+a^2E\sum_{i=1}^{2}\frac{1}{\wp'(\bar{x}_{i})}\left[\frac{(\lambda-\lambda_0)}{\mu}\zeta(\bar{x}_{i})+\ln{(\sigma(x-\bar{x}_i))}-\ln{(\sigma(x_0-\bar{x}_i))}\right]\label{Ithet} \ \ ,
\end{eqnarray}
where $\wp(\bar{x}_{1})=\wp(\bar{x}_{2})=-\nu/\mu$. The simple poles of the function (\ref{rfnWeierstrass}) given by $\bar{x}_1$, $\bar{x}_2$ are in the fundamental domain  $\{2a\omega_{1}+2b\omega_{2}|a,b\in [0,1]\}$ where $2\omega_{1}$ $\in$ $\mathbb{R}$ and $2\omega_{2}$ $\in$ $\mathbb{C}$.
In addition, we have
\begin{eqnarray}
\bar{I}_{r}=C_0(\lambda-\lambda_0)+\sum_{i=1}^{4}\sum_{j=1}^{2}\frac{ C_{i}}{\wp'(u_{ji})}\left[\frac{-(\lambda-\lambda_{0})}{\gamma}\zeta(u_{ji})+\ln{(\sigma(u-u_{ji}))}-\ln{(\sigma(u_{0}-u_{ji}))}\right]\label{Irt} \ \ ,
\end{eqnarray}
where $C_i$ are the coefficients from the partial fraction and and $\tilde{e}_i$ the poles of the rational function $F(z)$ from the transformation $(r^2+a^2)\Delta(r)^{-1}P(r)\frac{dr}{\sqrt{R(r)}}$. In addition $\wp(u_{1i})$ = $\wp(u_{2i})$ = $\tilde{e}_i$.

\section{Examples of orbits}
In the following we give plots of geodesics of massive and massless particles, respectively
in the space--time of a non--extremal Kerr black hole pierced by an infinitely thin cosmic string.
In particular we will demonstrate how the presence of the cosmic string
alters the test particle motion.
\subsection{Motion of massive test particles}
\begin{figure}[h!]
\begin{center}
\resizebox{4in}{!}{\includegraphics{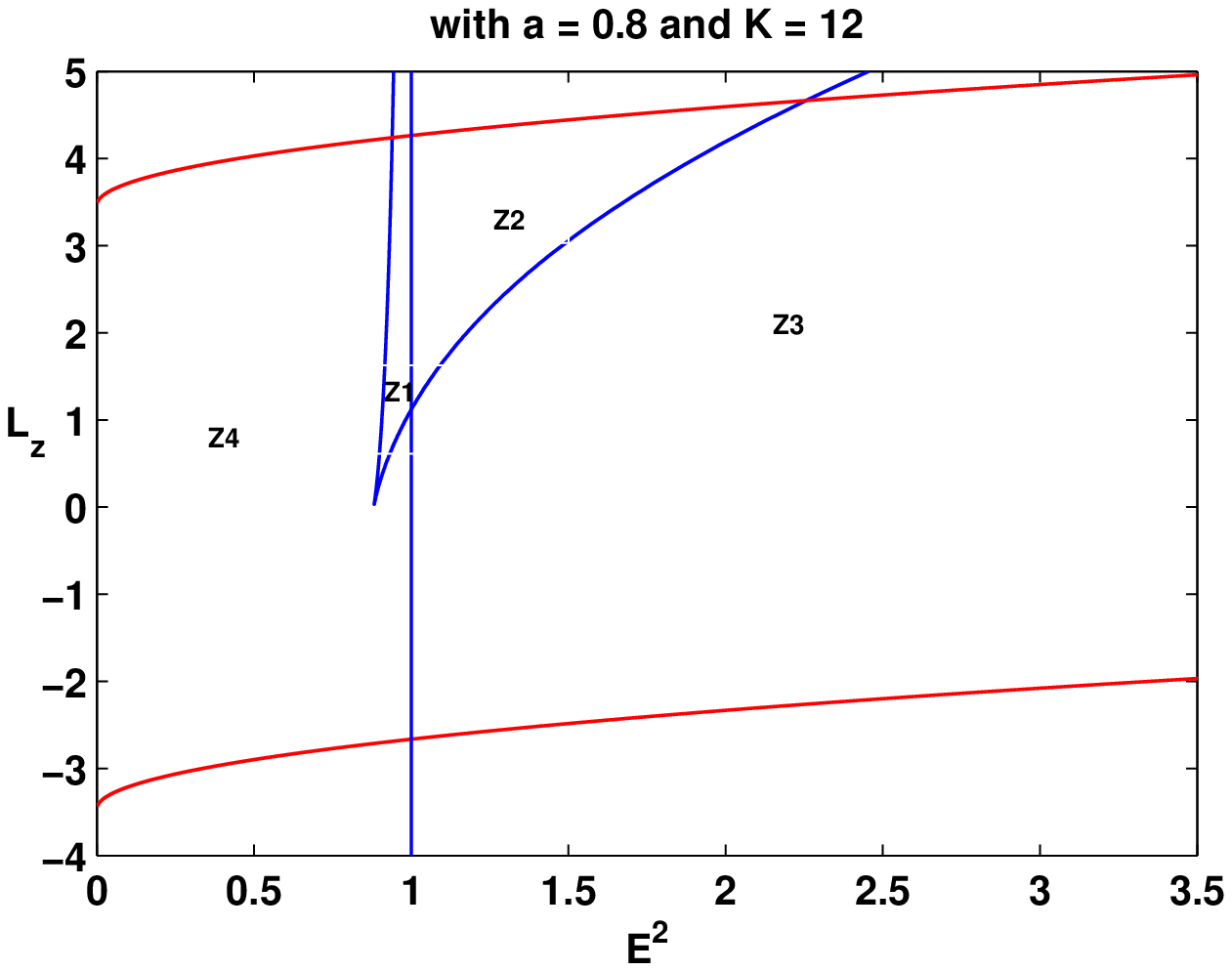}}
\end{center}
\caption{The four domains Z1 to Z4 in the  $L_z$-$E^2$--plane for which
solutions to the geodesic equation for massive particles exist are shown for $M=1$,  $a= 0.8$ and $K = 12$. The red and blues lines come from the restriction $\Theta(\theta) > 0$ and $R(r)>0$, respectively.}\label{za08K12}.
\end{figure}

The domain of existence of solutions of the geodesic equation 
can be obtain from the intersection of the allowed domains of the $L_z$-$E^2$-plane obtained
from the requirement $\Theta(\theta) >0$ and $R(r) >0$, respectively.
 This leads to four domains in the $L_z$-$E^2$-plane. These are shown
for $M=1$, $a=0.8$ and $K = 12$ in Fig.\ref{za08K12} and denoted by Z1 to Z4.

\begin{figure}[p!]
\centering
\subfigure[$\beta$ = 1]{
\includegraphics[scale=0.5]{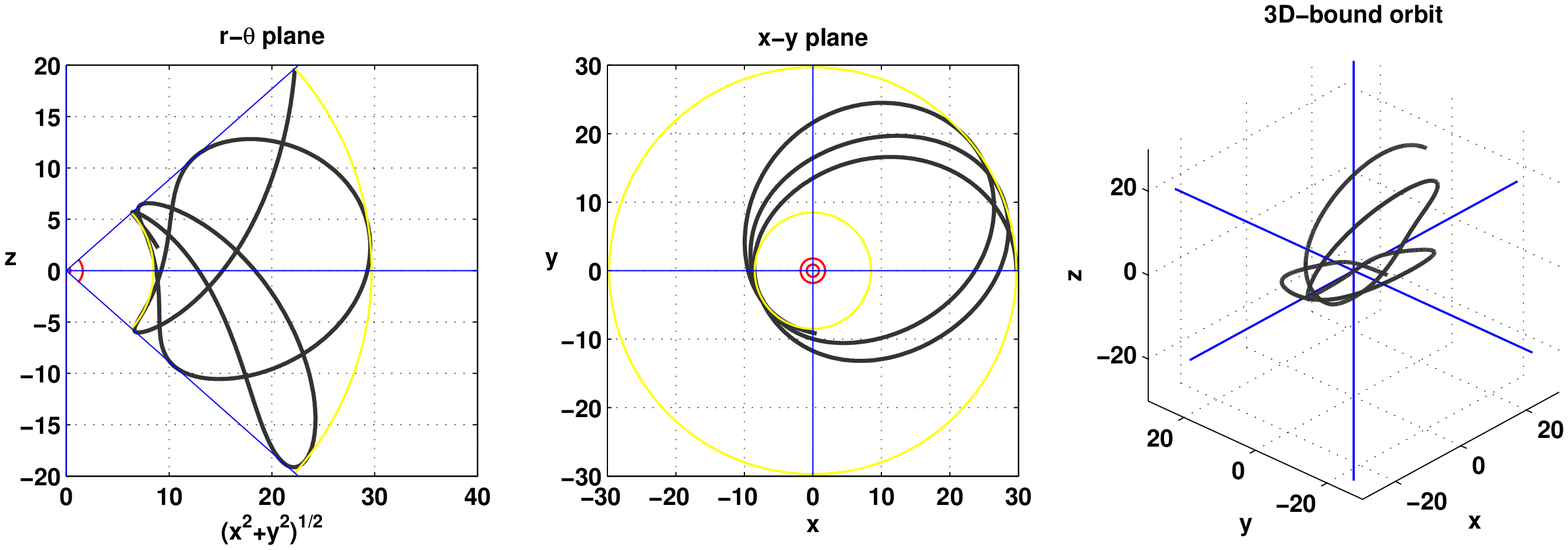}
\label{z2fbbeta1}
}

\subfigure[$\beta$ = 0.88]{
\includegraphics[scale=0.5]{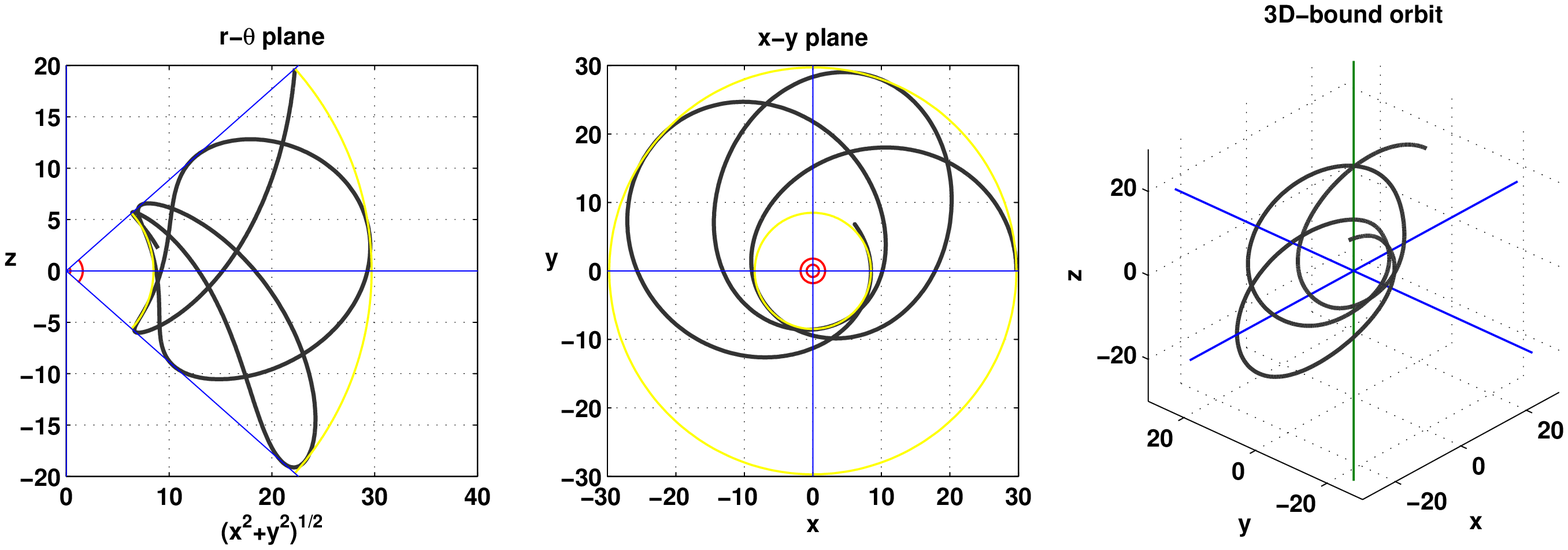}
\label{z2fbbeta075}
}

\subfigure[$\beta$ = 0.78]{
\includegraphics[scale=0.5]{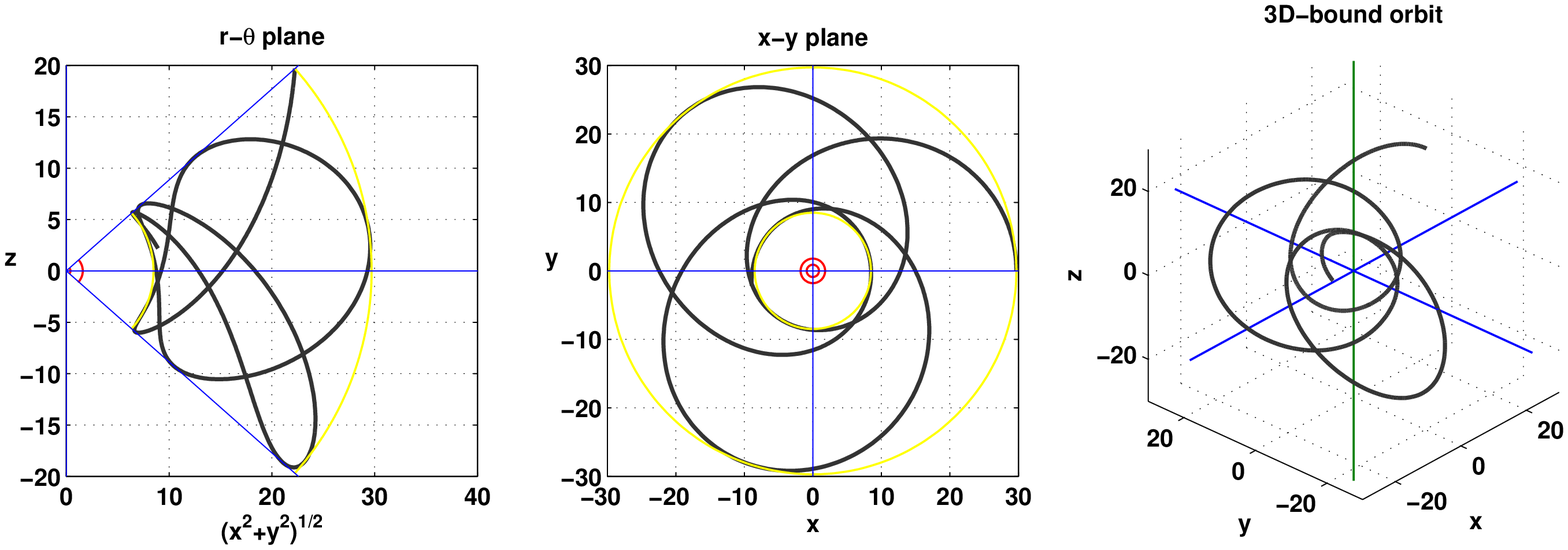}
\label{z2fbbeta075}
}

\caption[Optional caption for list of figures]{The change of the outer bound orbit of a massive test particle (domain Z1) due to the change of the deficit parameter $\beta$.
Here $L_z=3.0$, $E=\sqrt{0.95}$, $K =12$, $M =1$ and $a = 0.8$. The red circles represent the radii of the event and Cauchy horizons, while the yellow circles denote the minimal and the maximal radius of the orbit, respectively.}\label{z1boutter}
\end{figure}

\begin{figure}[p!]
\centering
\subfigure[$\beta$ = 1]{
\includegraphics[scale=0.5]{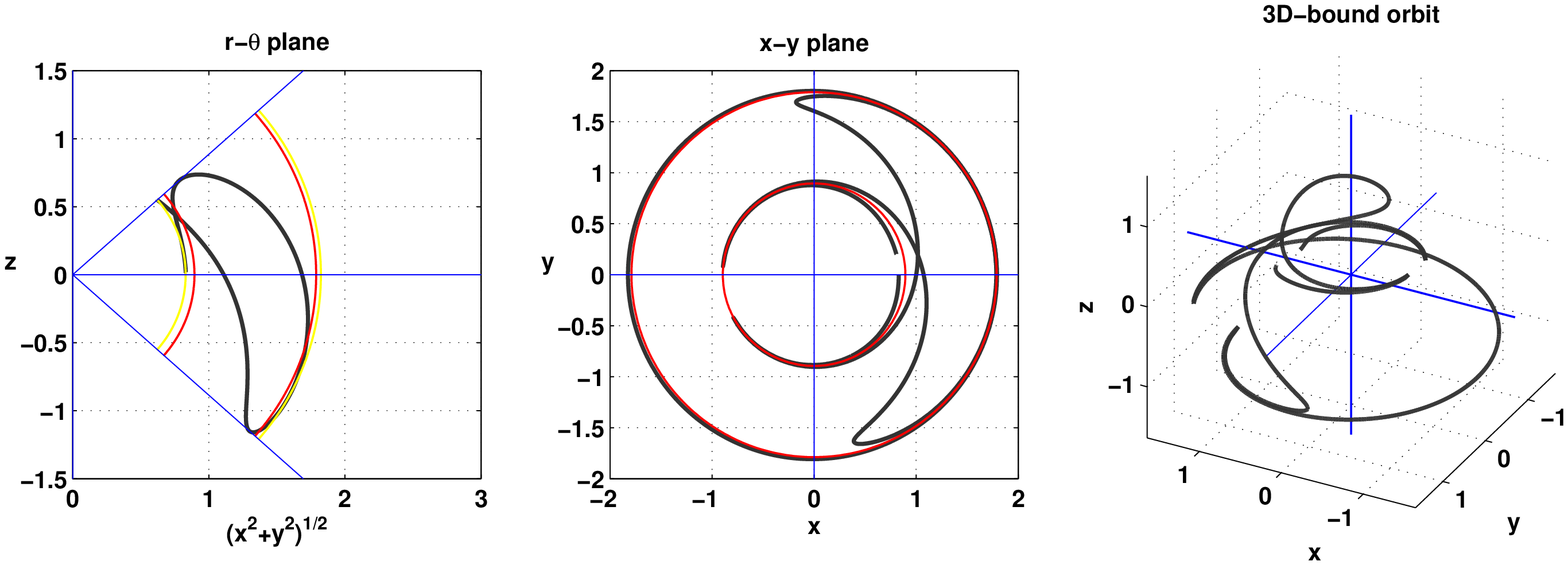}
\label{z2fbbeta1}
}

\subfigure[$\beta$ = 0.70]{
\includegraphics[scale=0.5]{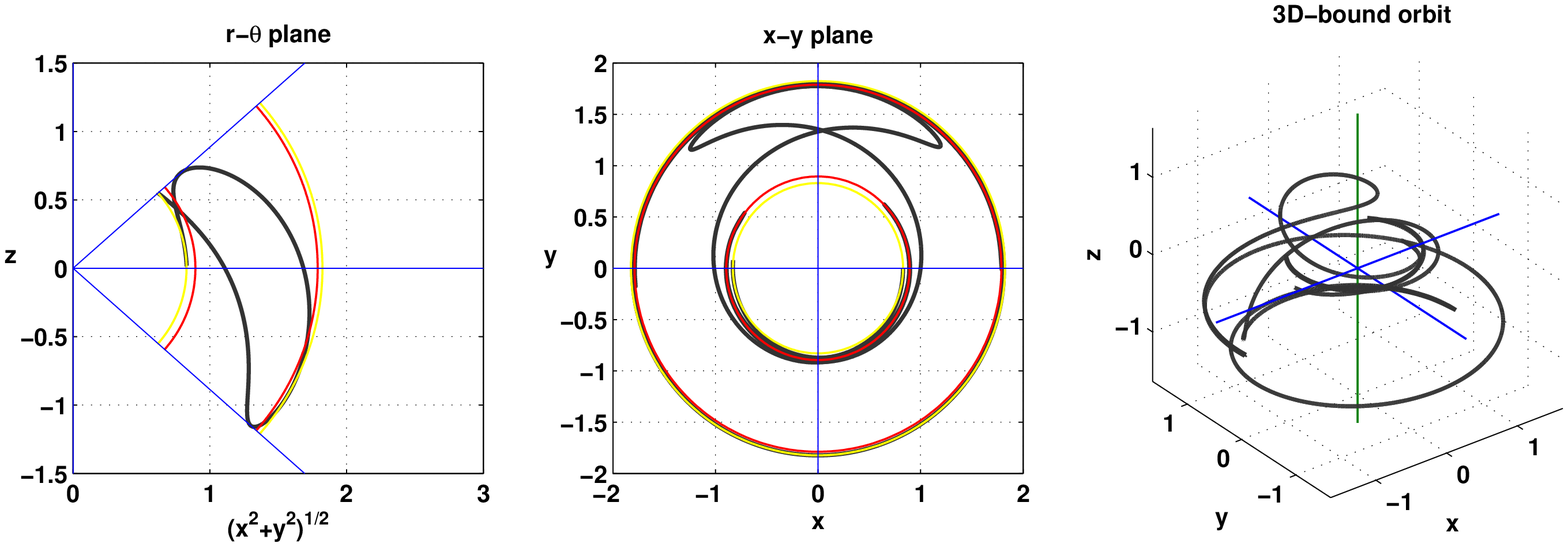}
\label{z2fbbeta075}
}

\subfigure[$\beta$ = 0.50]{
\includegraphics[scale=0.5]{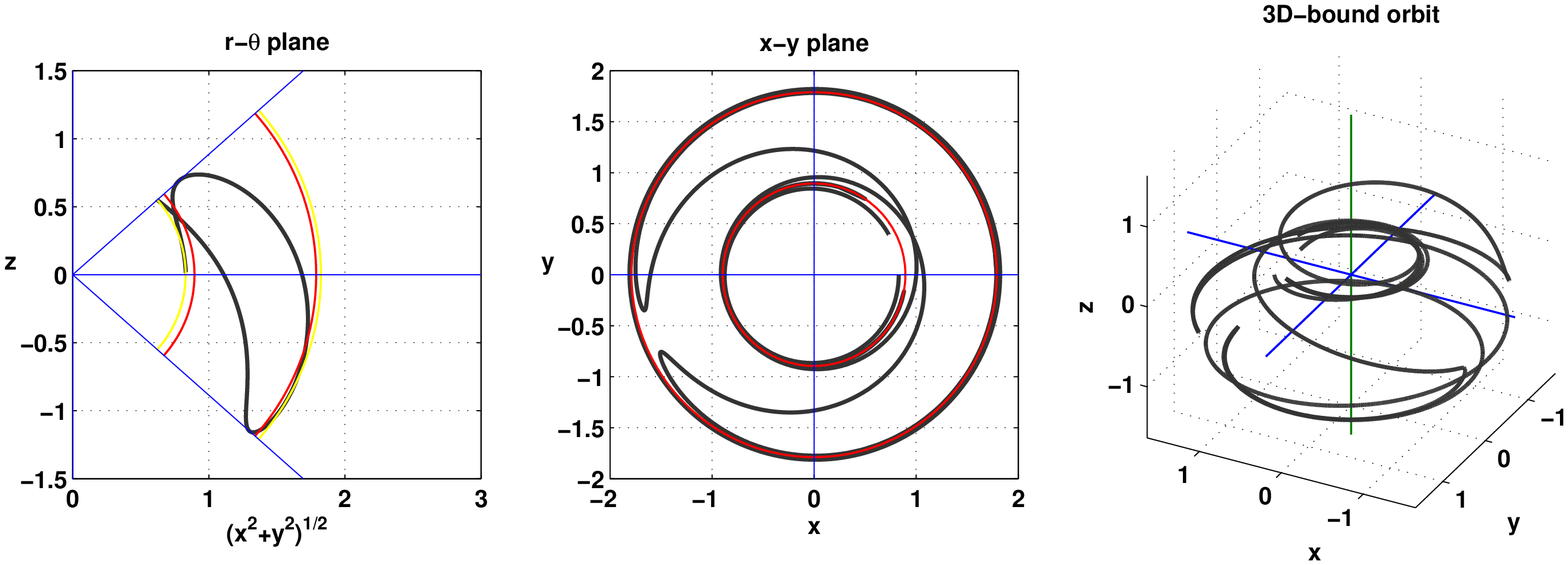}
\label{z2fbbeta075}
}

\caption[Optional caption for list of figures]{
The change of the inner bound orbit of a massive test particle (domain Z1) due to the change of the deficit parameter $\beta$.
Here $L_z=3.0$, $E=\sqrt{0.95}$, $K =12$, $M =1$ and $a = 0.8$. The red circles represent the radii of the event and Cauchy horizons, while the yellow circles denote the minimal and the maximal radius of the orbit, respectively.}\label{z1innerbound}
\end{figure}

\begin{figure}[p!]
\centering
\subfigure[$\beta$ = 1]{
\includegraphics[scale=0.5]{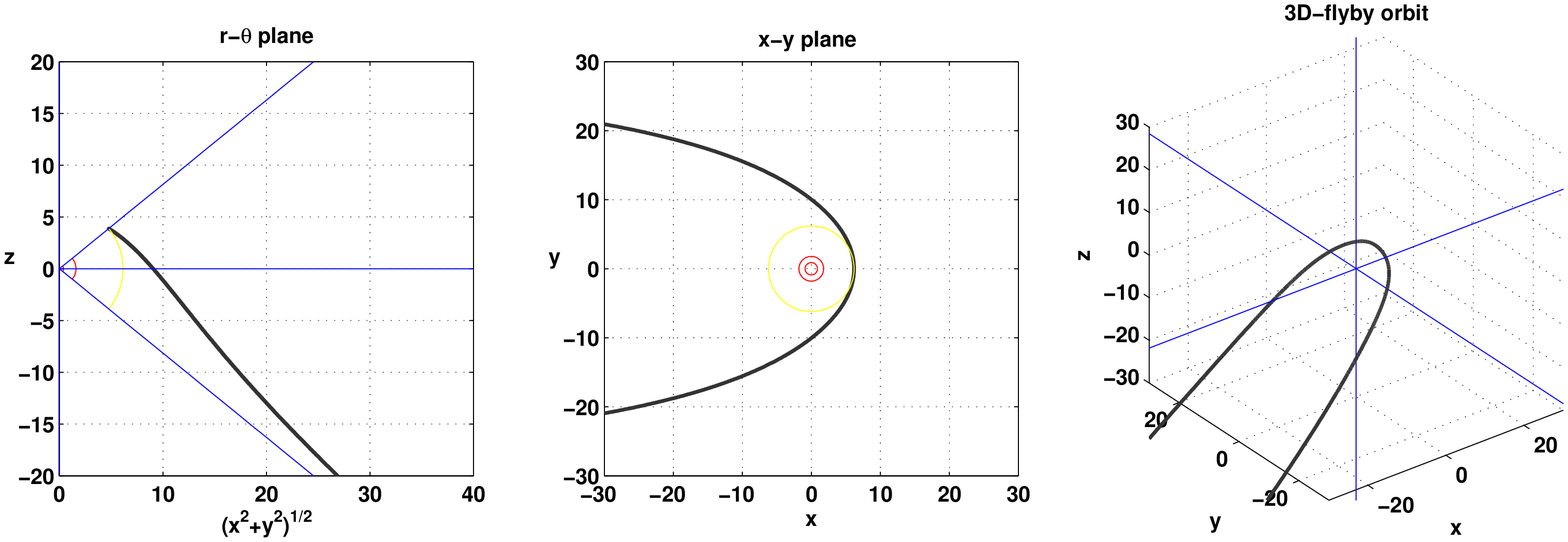}
\label{z2fbbeta1}
}

\subfigure[$\beta$ = 0.60]{
\includegraphics[scale=0.5]{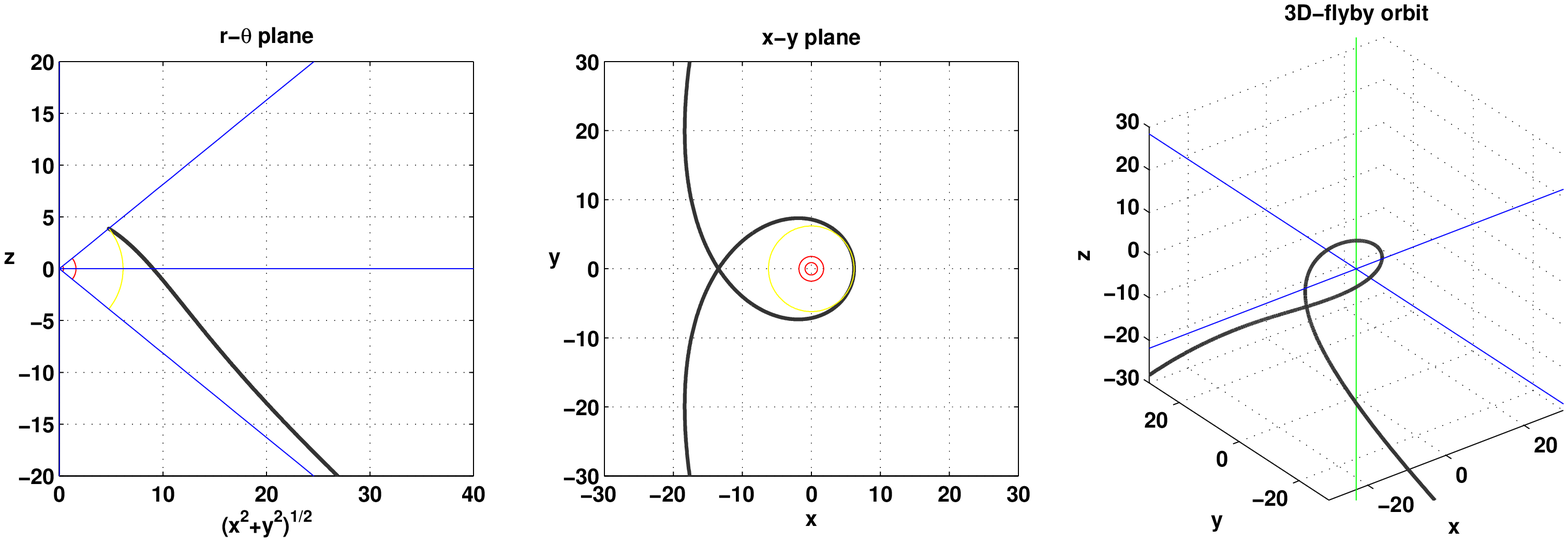}
\label{z2fbbeta075}
}

\subfigure[$\beta$ = 0.50]{
\includegraphics[scale=0.5]{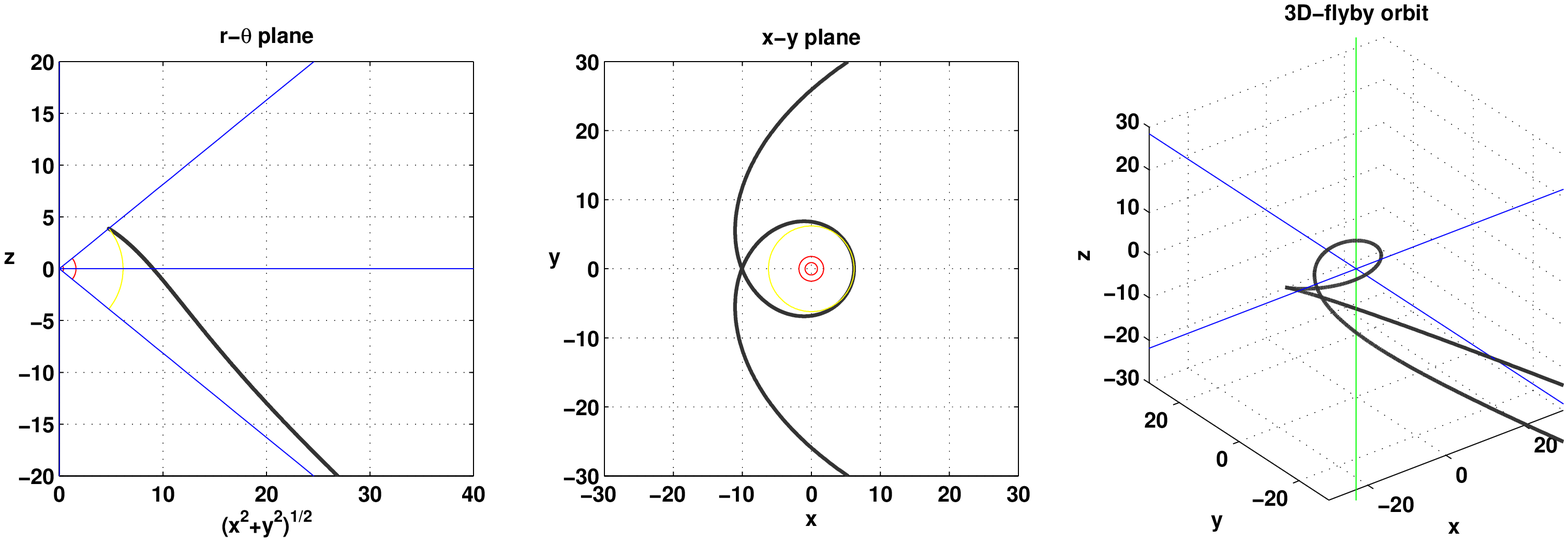}
\label{z2fbbeta075}
}

\caption[Optional caption for list of figures]{
The change of the flyby orbit (domain Z2) of a massive test particle due to the change of the deficit parameter $\beta$. Here $L_z=3.104$, $E=1.004$, $K =12$, $M =1$ and $a = 0.8$. The red circles represent the radii of the event and Cauchy horizons, while the yellow circle denotes the minimal radius of the orbit.}\label{z2fb}
\end{figure}

\begin{figure}[p!]
\centering
\subfigure[$\beta$ = 1]{
\includegraphics[scale=0.5]{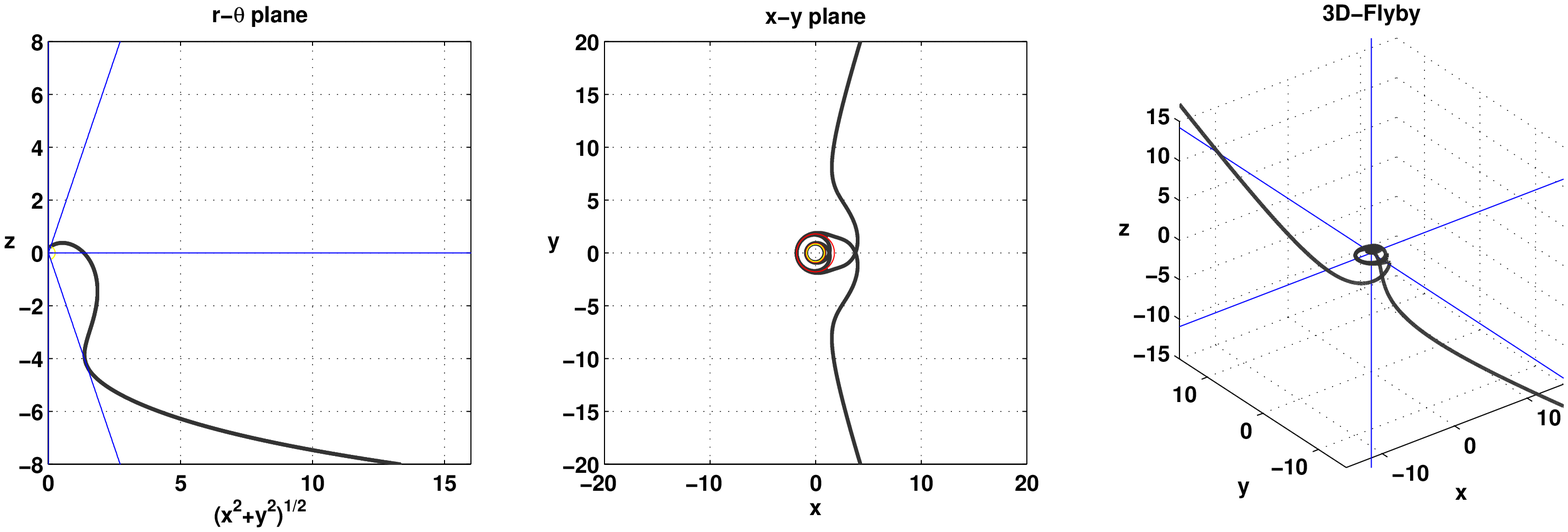}
\label{z2fbbeta1}
}

\subfigure[$\beta$ = 0.65]{
\includegraphics[scale=0.5]{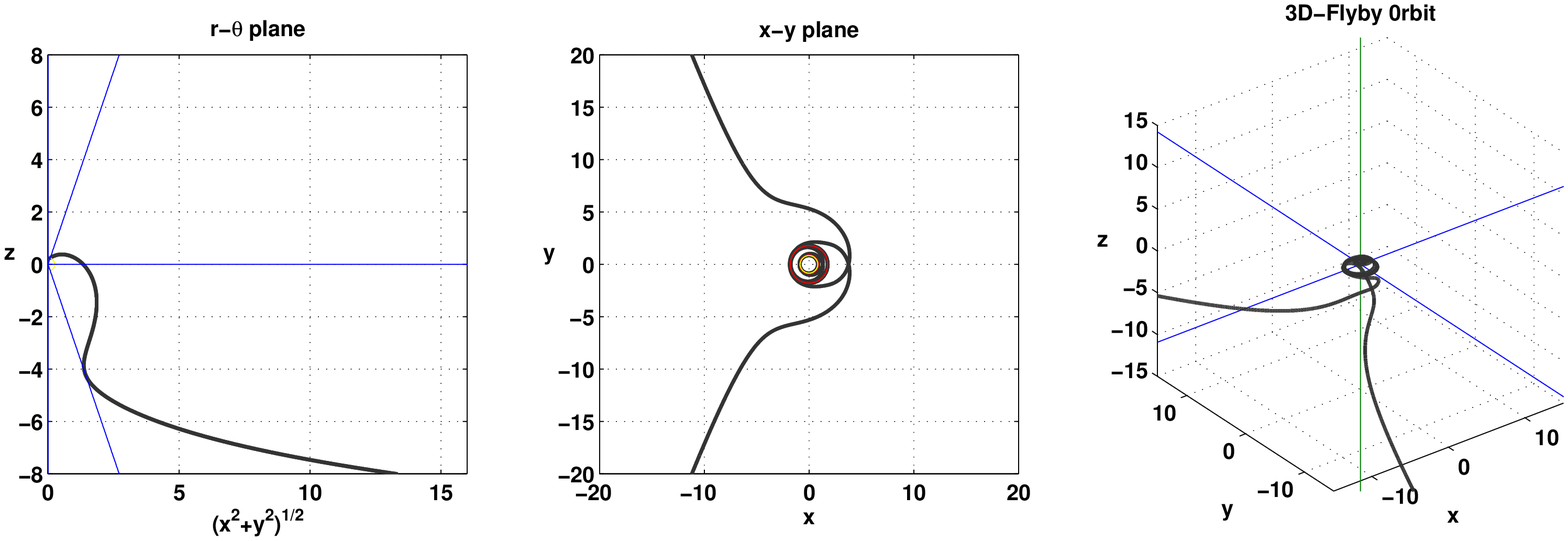}
\label{z2fbbeta075}
}

\subfigure[$\beta$ = 0.50]{
\includegraphics[scale=0.5]{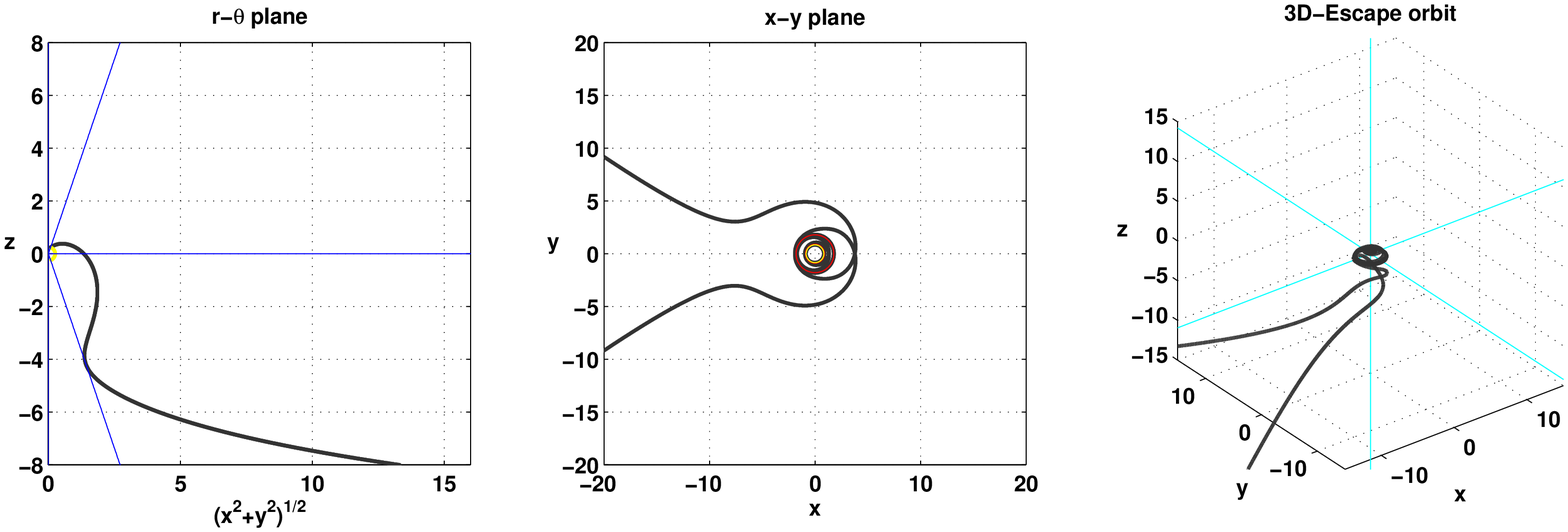}
\label{z2fbbeta075}
}

\caption[Optional caption for list of figures]{
The change of the flyby orbit (domain Z3) of a massive test particle due to the change of the deficit parameter $\beta$. Here $L_z=-1$, $E=\sqrt{1.10}$, $K =12$, $M =1$ and $a = 0.8$. The red circles represent the radii of the event and Cauchy horizons, while the yellow circle denotes the minimal radius of the orbit.}\label{fbz3}
\end{figure}

\begin{figure}[p!]
\centering
\subfigure[$\beta$ = 1]{
\includegraphics[scale=0.5]{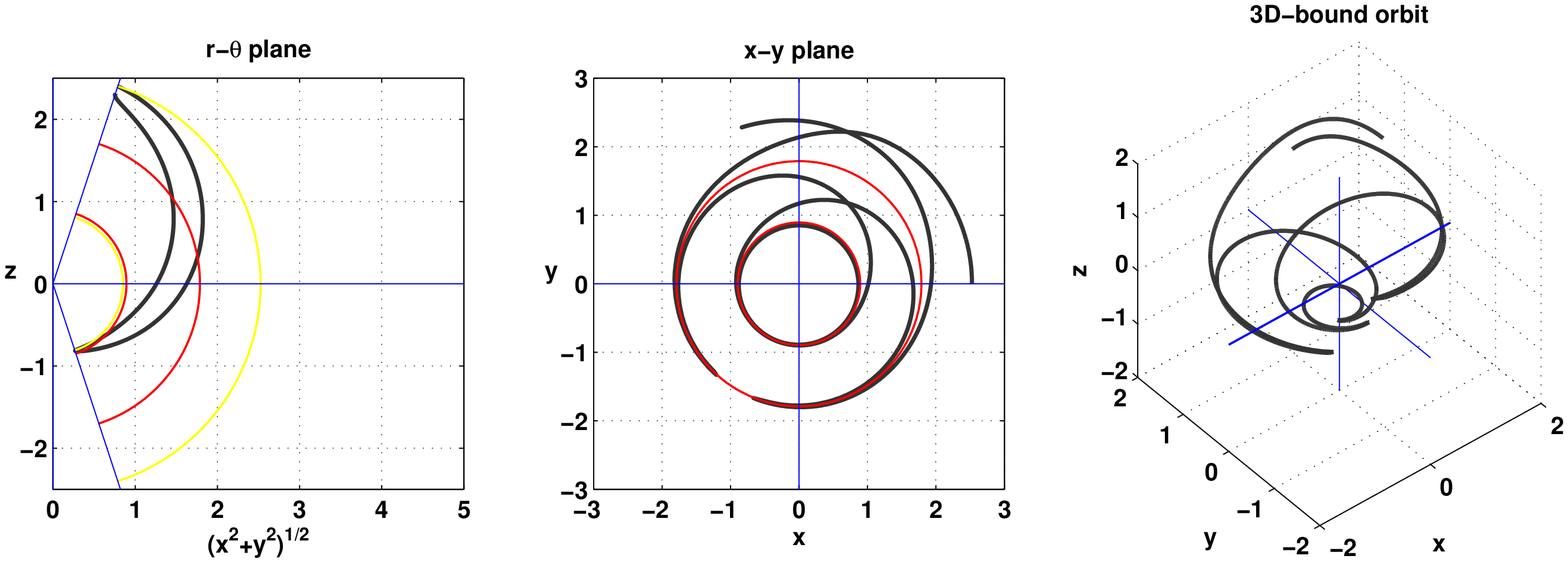}
\label{z2fbbeta1}
}

\subfigure[$\beta$ = 0.85]{
\includegraphics[scale=0.5]{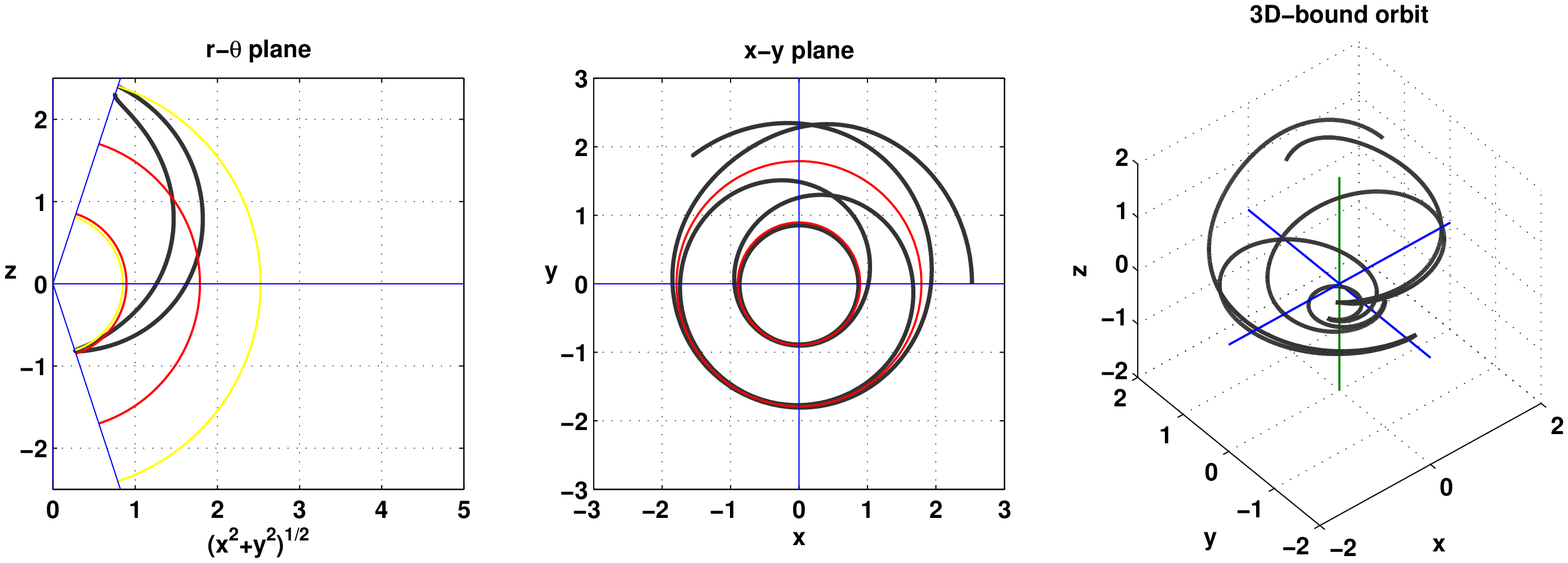}
\label{z2fbbeta085}
}

\subfigure[$\beta$ = 0.70]{
\includegraphics[scale=0.5]{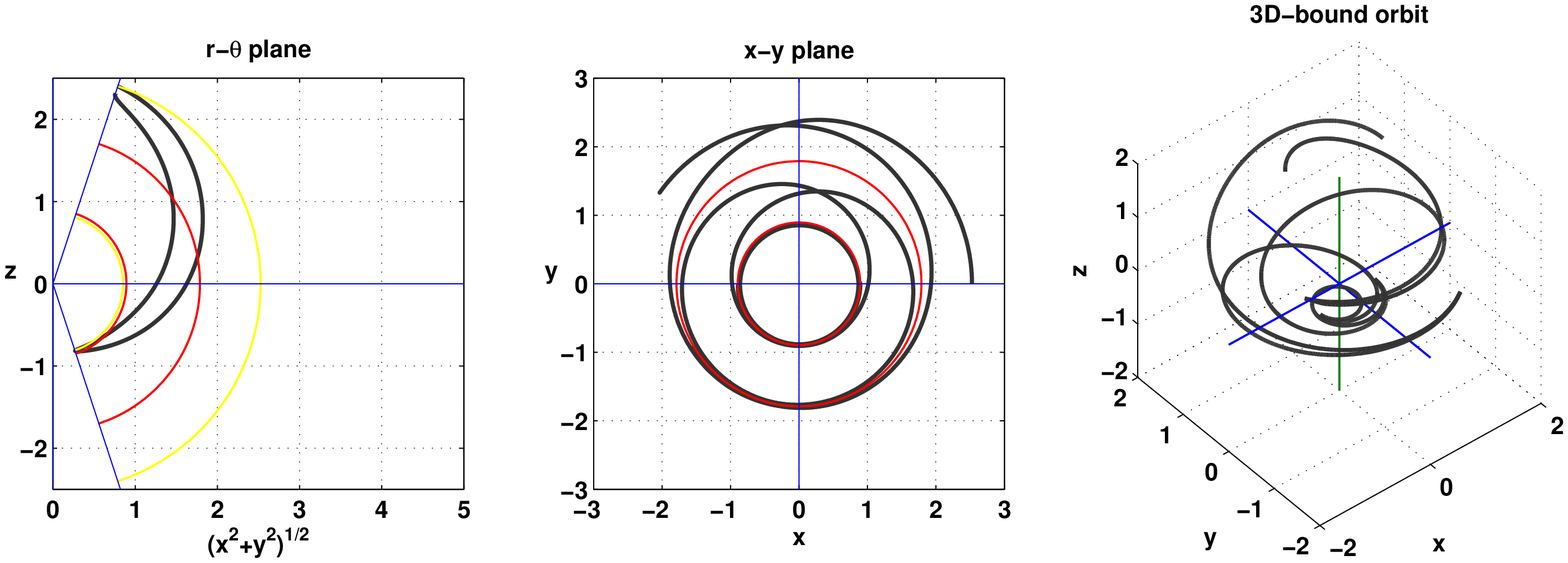}
\label{z2fbbeta085}
}

\caption[Optional caption for list of figures]{
The change of the bound orbit (domain Z4) of a massive test particle due to the change of the deficit parameter $\beta$. Here $L_z=-1$, $E=\sqrt{0.5}$, $K =12$, $M =1$ and $a = 0.8$. The red circles represent the radii of the event and Cauchy horizons, while the yellow circles denote the minimal and maximal radius of the orbit, respectively.}
\label{ibz4}
\end{figure}

\begin{enumerate}
 \item \textbf{Z1:} This domain is the combination of domain $T4$ from Fig.\ref{LzET} and domain $R3$ from Fig.\ref{LzER}. 
 The possible orbits are two bound orbits on which the test particle can cross the equatorial plane at $\theta$ = $\pi$/2. The effect of the cosmic string on the geodesic
motion on the outer bound and inner bound orbit is shown in Fig.\ref{z1boutter} and Fig.\ref{z1innerbound}, respectively. Note that the test particle on the inner bound orbit (see Fig.\ref{z1innerbound}) crosses the event and the Cauchy horizon (red circles) several times.

 \item \textbf{Z2:} This domain is the combination of domain $T4$ from Fig.\ref{LzET} and domain $R4$ from Fig.\ref{LzER}. 
The possible orbits are one flyby and one bound orbit on which the test particle can cross the equatorial plane at $\theta$ = $\pi$/2.  The effect of the cosmic string on the flyby orbit is shown in Fig.\ref{z2fb}.

 \item \textbf{Z3:} This domain is the combination of domain $T4$ from Fig.\ref{LzET} and domain $R2$ from Fig.\ref{LzER}. 

 The possible orbits are two flyby orbits (from $r_1$ to $\infty$ and $-\infty$ to $r_2$) on which the test particle can cross the equatorial plane at $\theta$ = $\pi$/2. The effect of
the cosmic string on the flyby orbit (from $r_1$ to $\infty$) is shown in Fig.\ref{fbz3}. Note that the test particle crosses the event and the Cauchy horizon (red circles) several times.

 \item \textbf{Z4:} This domain is the combination of domain $T4$ from Fig.\ref{LzET} and domain $R5$ from Fig.\ref{LzER}.

 The possible orbit is one bound orbit  on which the test particle can cross the equatorial plane at $\theta$ = $\pi$/2. The effect of the cosmic string on this bound orbit is shown in Fig.\ref{ibz4}. Note that the test particle crosses the event and the Cauchy horizon (red circles) several times.
\end{enumerate}

\subsection{Motion of massless test particles}
\begin{figure}[p!]
\begin{center}
\resizebox{6.0in}{!}{\includegraphics{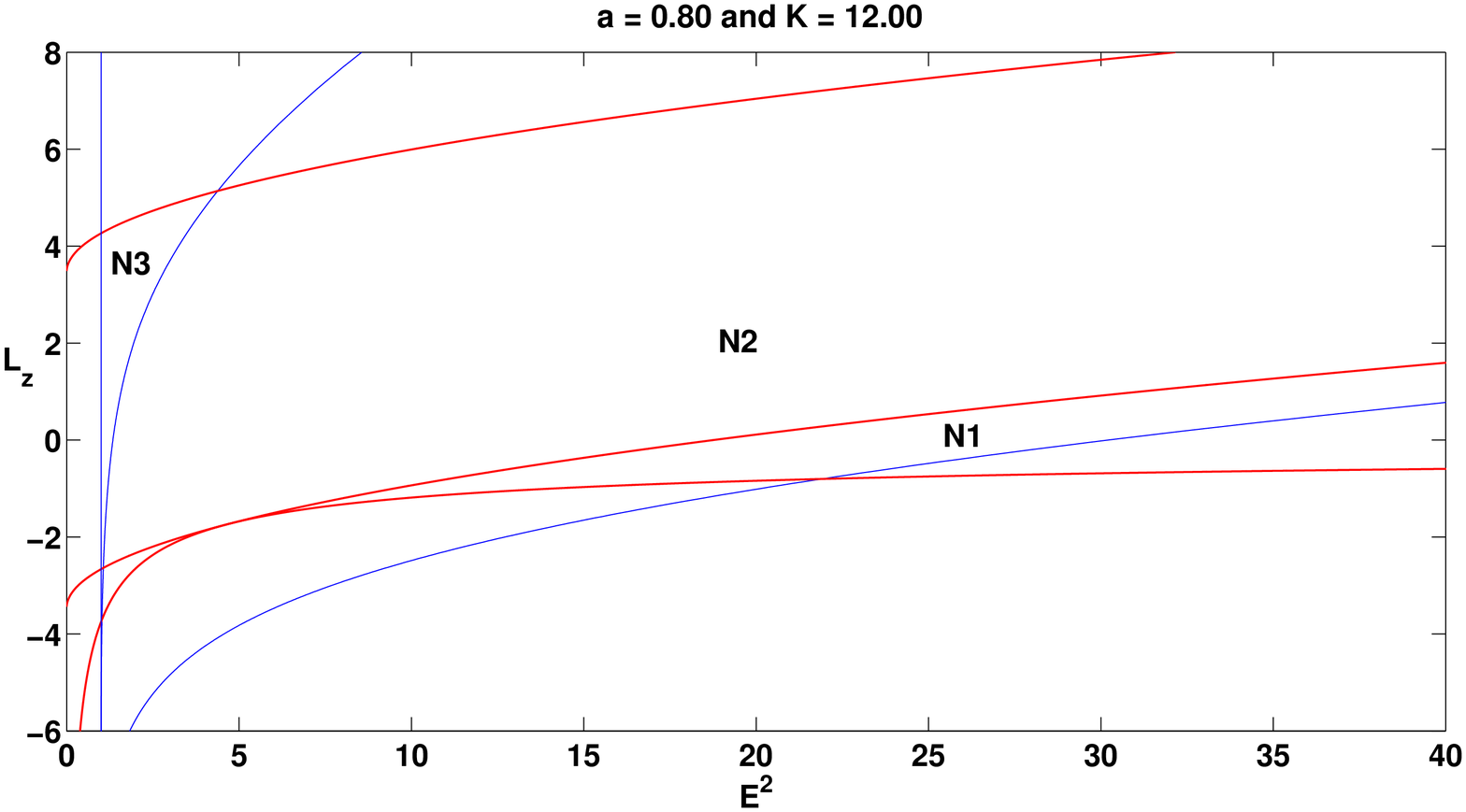}}
\end{center}
\caption{ 
The three domains N1 to N3 in the  $L_z$-$E^2$--plane for which
solutions to the geodesic equation for massless particles exist are shown for $M=1$,  $a= 0.8$ and $K = 12$. The red and blues lines come from the restriction $\Theta(\theta) > 0$ and $R(r)>0$, respectively.}\label{za08K12null}
\end{figure}

The domain of existence of solutions of the geodesic equation 
can be obtain from the intersection of the allowed domains of the $L_z$-$E^2$-plane obtained
from the requirement $\Theta(\theta) >0$ and $R(r) >0$, respectively.
This leads to three domains in the $L_z$-$E^2$-plane. These are shown
for $M=1$, $a=0.8$ and $K = 12$ in Fig. \ref{za08K12null} and denoted by N1 to N3.

\begin{figure}[p!]
\centering
\subfigure[$\beta$ = 1]{
\includegraphics[scale=0.5]{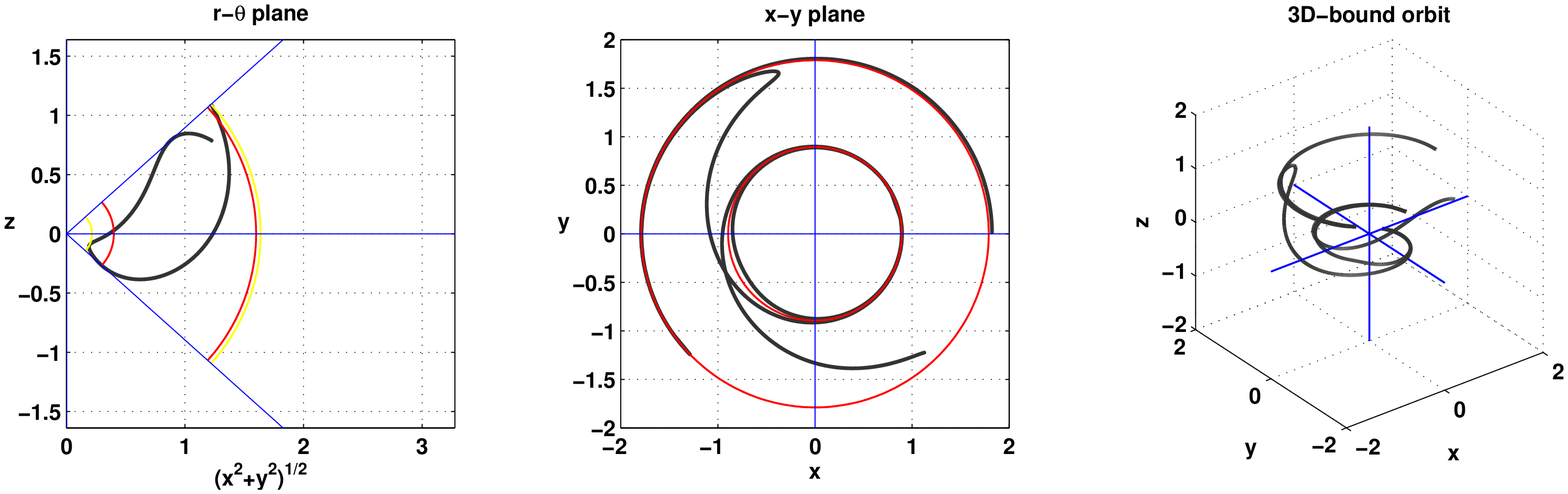}
\label{z2fbbeta1}
}

\subfigure[$\beta$ = 0.75]{
\includegraphics[scale=0.5]{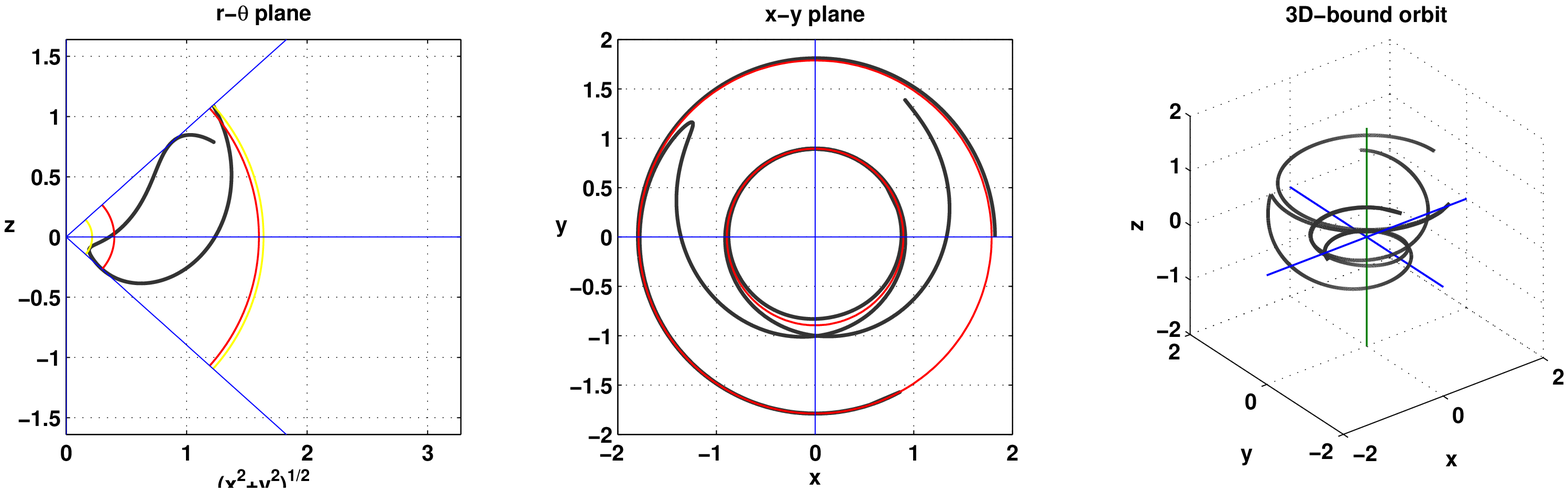}
\label{z2fbbeta085}
}

\subfigure[$\beta$ = 0.60]{
\includegraphics[scale=0.5]{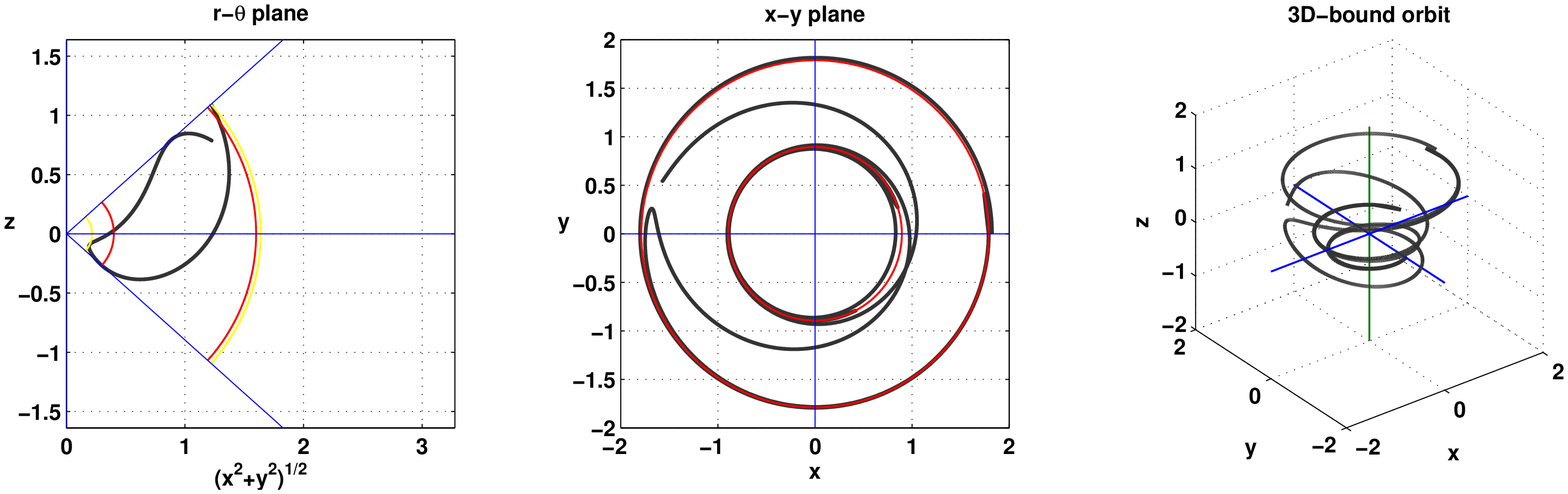}
\label{z2fbbeta085}
}

\caption[Optional caption for list of figures]{
The change of the bound orbit (domain N3) of a massless test particle due to the change of the deficit parameter $\beta$. Here $L_z=3.0$, $E=\sqrt{0.90}$, $K =12$, $M =1$ and $a = 0.8$. The red circles represent the radii of the event and Cauchy horizons, while the yellow circles denote the minimal and maximal radius of the orbit, respectively.}\label{n3inbmlz5}
\end{figure}

\begin{figure}[p!]
\centering
\subfigure[$\beta$ = 1]{
\includegraphics[scale=0.5]{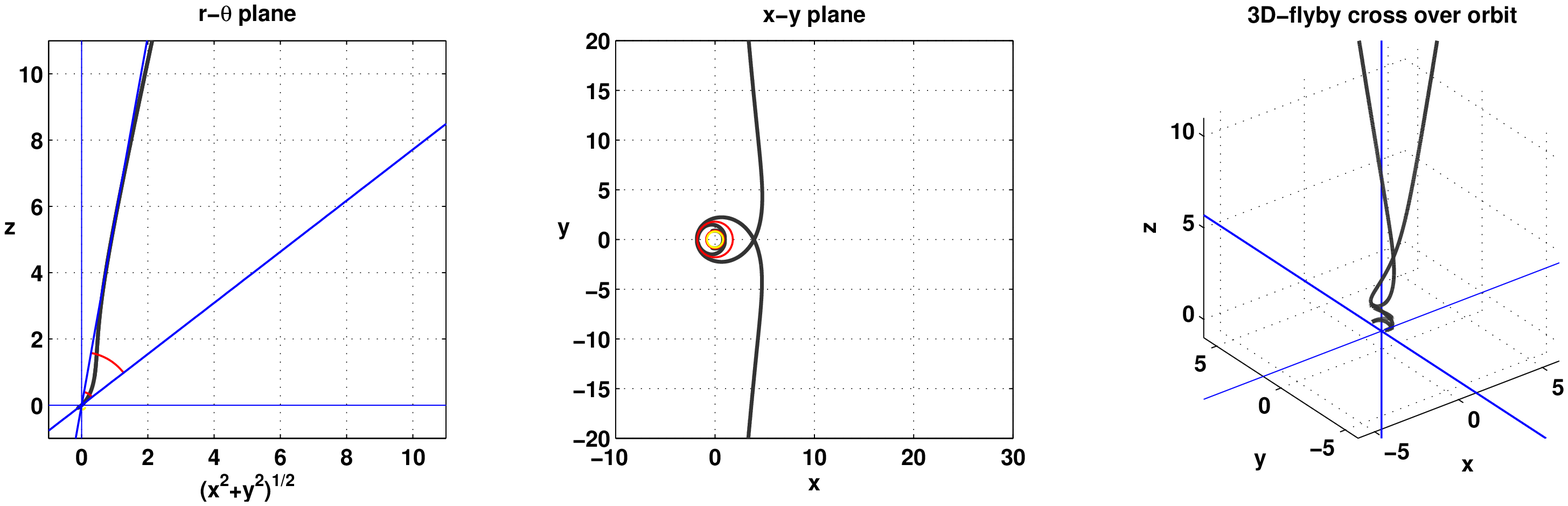}

}

\subfigure[$\beta$ = 0.80]{
\includegraphics[scale=0.5]{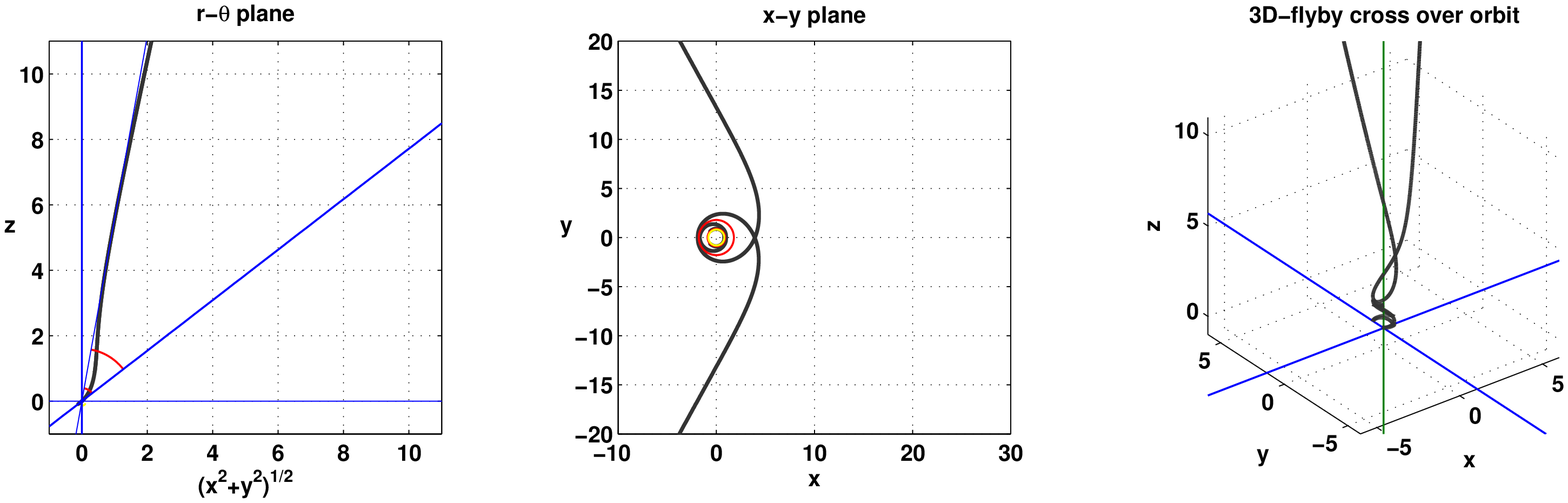}

}

\subfigure[$\beta$ = 0.20]{
\includegraphics[scale=0.5]{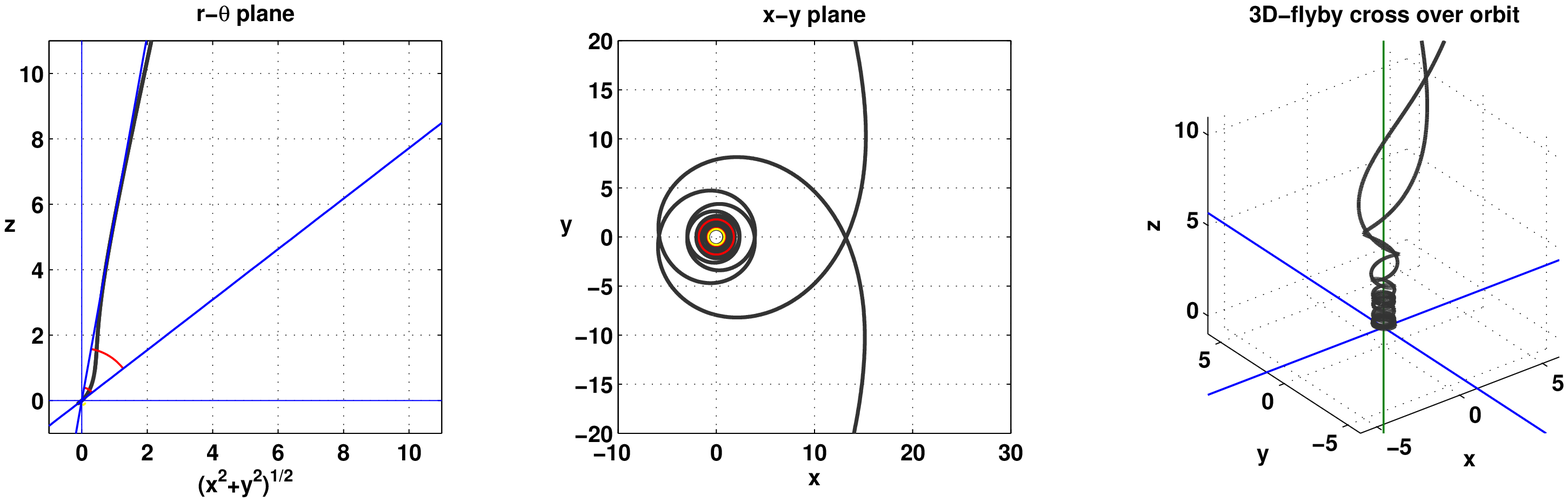}

}

\caption[Optional caption for list of figures]{
The change of the cross--over flyby orbit (domain N1) of a massless test particle due to the change of the deficit parameter $\beta$. Here $L_z=-0.50$, $E=\sqrt{20}$, $K =12$, $M =1$ and $a = 0.8$. The red circles represent the radii of the event and Cauchy horizons, while the yellow circle denotes the minimal radius of the orbit.}\label{n2fbcmlz5}
\end{figure}

\begin{figure}[p!]
\centering
\subfigure[$\beta$ = 1]{
\includegraphics[scale=0.5]{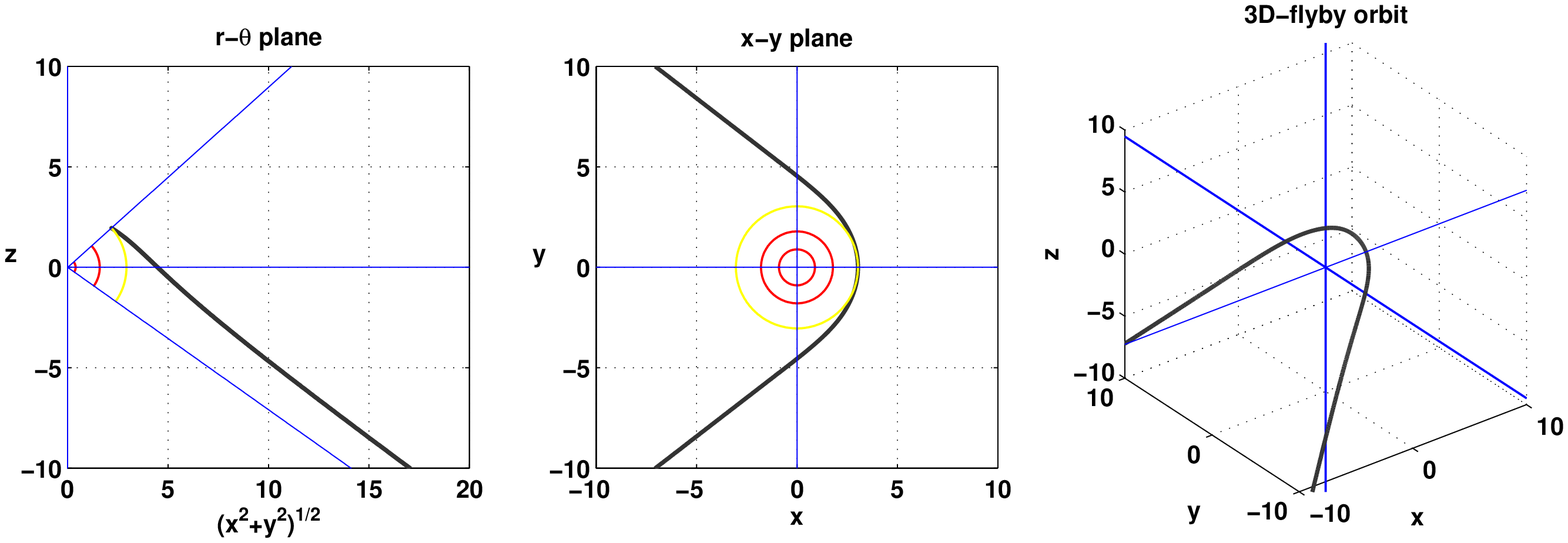}
\label{z2fbbeta1}
}

\subfigure[$\beta$ = 0.50]{
\includegraphics[scale=0.5]{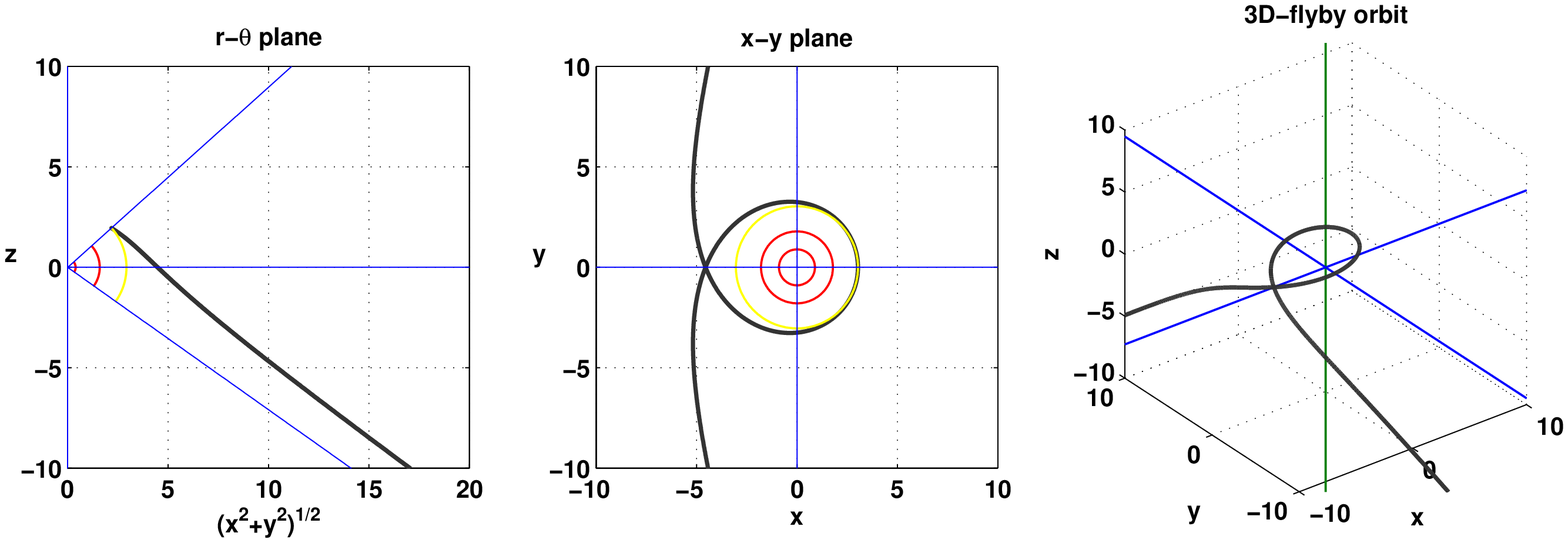}
\label{z2fbbeta085}
}

\subfigure[$\beta$ = 0.25]{
\includegraphics[scale=0.5]{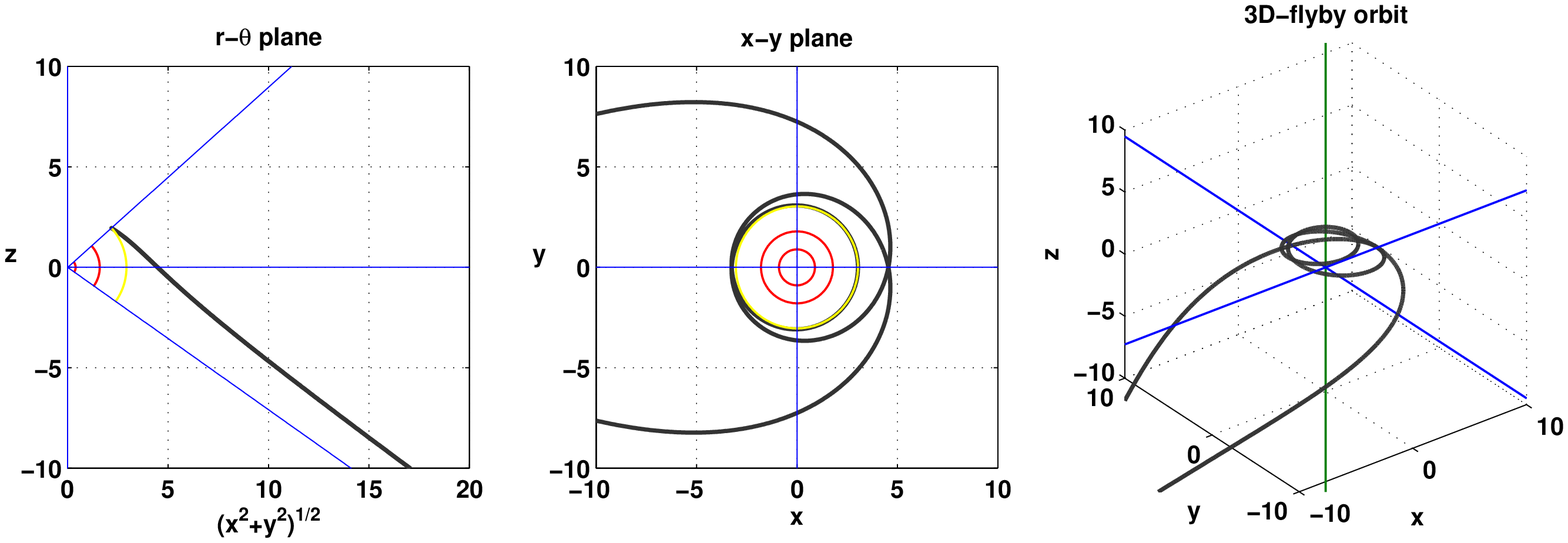}
\label{z2fbbeta085}
}

\caption[Optional caption for list of figures]{
The change of the flyby orbit (domain N3) of a massless test particle due to the change of the deficit parameter $\beta$. Here $L_z=3$, $E=\sqrt{0.9}$, $K =12$, $M =1$ and $a = 0.8$. The red circles represent the radii of the event and Cauchy horizons, while the yellow circle denotes the minimal radius of the orbit.}\label{n3fbmlz5}
\end{figure}

\begin{enumerate}
 \item \textbf{N1}: The possible orbit is a  cross--over flyby orbit on which the test particle cannot cross the equatorial plane at $\theta$ = $\pi$/2.
 The effect of
the cosmic string on the cross--over flyby orbit (from $r_1$ to $\infty$) is shown in Fig.\ref{n2fbcmlz5}

 \item \textbf{N2}: The possible orbits are one bound and one flyby orbit on which the test particle can cross the equatorial plane at $\theta$ = $\pi$/2. 
The effect of
the cosmic string on the cross--over flyby orbit (from $r_1$ to $\infty$) is shown in Fig.\ref{z2fb}.

 \item \textbf{N3}: The possible orbits are one flyby and one bound orbit on which the test particle can cross  the equatorial plane at $\theta$ = $\pi$/2.
The effect of
the cosmic string on the flyby and on the bound orbit is shown
    in Fig.\ref{n3fbmlz5} and Fig.\ref{n3inbmlz5}, respectively.

\end{enumerate}

\section{Observables}

For the standard Kerr space--time ($\beta=1$) the analytical expression of the so--called fundamental frequencies 
of bound orbits have been given in  \cite{DrascoHughes2004} in the frequency domain using a Fourier transformation and in \cite{Fujita2009} as a function of a Jacobi elliptic integral. Here, we will present the fundamental frequencies
for $\beta\le 1$  in the  form of a Weierstrass elliptic integral $\wp$ and analyze their
dependence on the deficit angle.

For bound orbits $r(\lambda)$ and $\theta(\lambda)$ become periodic functions that are independent of each other.
We can then define the fundamental frequencies of these bound orbits using Mino time $\lambda$. Using the
results of Sections  \ref{4r2r2c} and \ref{thetasection} we find
\begin{eqnarray}
\Lambda_r&=&2\int_{r_{min}}^{r_{max}}\frac{dr}{\sqrt{R(r)}}=2\gamma\tilde{\omega_1} \ \ , \ \ \Lambda_{\theta}=4\int_{0}^{\cos\theta_{min}}\frac{d\cos \theta}{\sqrt{\Theta(\cos\theta)}}=4\mu\omega_{1}  \ ,
\end{eqnarray}
where $r(\lambda)$ = $r(\lambda+n\Lambda_r)$ and $\theta(\lambda)$ = $\theta(\lambda+n\Lambda_{\theta})$ for any integer $n$ \cite{DrascoHughes2004}. $r_{\rm min}$ and $r_{\rm max}$ correspond to the periapsis and the
apoapsis in the radial direction, while $\theta_{\rm min}$ corresponds to the minimal value of the
polar coordinate. 
The angular frequencies then read
\begin{eqnarray}
\Upsilon_r=\frac{2\pi}{\Lambda_r}=\frac{\pi}{\gamma\omega_{r1}}\ \ \ , \ \ \Upsilon_{\theta}=\frac{2\pi}{\Lambda_{\theta}}=\frac{\pi}{2\mu\omega_{1}}  \ .
\end{eqnarray}
In \cite{Fujita2009} the frequencies for the $\phi$-- and $t$--component have been
defined via an average over the orbital periods $\Lambda_r$ and $\Lambda_{\theta}$. Here we read these
periods off from our solutions for the $\phi$-- and $t$--component of the geodesic equation. These solutions contain terms that correspond to oscillations with periods $\Lambda_r$ and $\Lambda_{\theta}$, respectively and a term that describes a linear increase in Mino time $\lambda$. $\Upsilon_{\phi}$ and $\Gamma$ are the frequencies of $\phi$ and $t$ in Mino time,
respectively that correspond to this linear increase \cite{Fujita2009}.
By using the results of sections \ref{phimotion} and \ref{tmotion}, we can find the expressions for these frequencies. This gives
\begin{eqnarray}
\label{upsilon}
\Upsilon_{\phi}&=&\frac{1}{\beta}\left((L_{z}-aE-K_0)+\sum_{i,j=1}^2\left[\frac{\nu+\mu d_{0}}{\wp'(x_{i})}\frac{\zeta(x_{i})}{\mu}-\frac{ K_{j}}{\wp'(u_{ji})}\frac{\zeta(u_{ji})}{\gamma}\right)\right] \ \ \label{Uphi}
\end{eqnarray}
and
\begin{eqnarray}
\Gamma&=&a(L_z-aE)+C_0+\sum_{i=1}^{4}\sum_{j=1}^{2}\left[a^2E\frac{1}{\wp'(\bar{x}_{i})}\frac{\zeta(\bar{x}_{i})}{\mu}-\frac{ C_{i}}{\wp'(u_{ji})}\frac{\zeta(u_{ji})}{\gamma}\right]  \ .
\end{eqnarray}
These results differ from that of the standard Kerr space--time by the factor of $1/\beta$ in (\ref{upsilon}).

As shown in \cite{DrascoHughes2004} the angular frequencies calculated using Mino time $\lambda$ are related to the angular frequencies $\Omega_r$, $\Omega_{\theta}$ and $\Omega_{\phi}$ calculated using a distant observer time as follows
\begin{eqnarray}
\Omega_r=\frac{\Upsilon_r}{\Gamma} \ \ , \ \  \Omega_{\theta}=\frac{\Upsilon_{\theta}}{\Gamma} \ \ , \ \ \Omega_{\phi}=\frac{\Upsilon_{\phi}}{\Gamma}\label{gamma}  \ .
\end{eqnarray}

If these frequencies are different they give rise to the precession of the orbital ellipse
and of the orbital plane. In particular, the
perihelion shift is related to the difference between the angular frequency of the radial motion $\Omega_r$ and
the angular frequency of the $\phi$-motion, while the Lense--Thirring (LT) precession is related to the difference
between the frequencies of the two angular motions
\begin{equation}
 \Omega_{\rm perihelion}=\Omega_{\phi} -\Omega_{r} \ \ , \ \ \Omega_{\rm LT} = \Omega_{\phi}-\Omega_{\theta} \ .
\end{equation}
In comparison to the standard Kerr space--time with $\beta=1$, the frequencies of the perihelion
shift $\Omega_{\rm perihelion}$ and the Lense--Thirring precession $\Omega_{\rm LT}$ are hence
bigger for $\beta < 1$, i.e. in the Kerr space--time including a deficit angle.
This is e.g. clearly seen in Fig.~9, where the perihelion shift increases
for decreasing $\beta$ (see plots in the $x$-$y$-plane) and the increase of precession is
seen when studying the 3-d orbits.

Using the results from the LAGEOS satellites \cite{ciufolini}
we can estimate an upper bound for the deficit parameter and hence for the energy per unit length
of a cosmic string piercing the earth. The theoretical value of the $\Omega_{\rm LT}$ for the earth is given by $39\cdot 10^{-3}$ arcseconds/year.
The LAGEOS satellites have measured this value with an accuracy of 10$\%$. Using this we find that
\begin{equation}
\Omega_{\rm LT}(\beta\neq 1) - \Omega_{\rm LT}(\beta=1)=\left(\frac{1}{\beta}-1\right)\Omega_{\phi}(\beta=1) \le 4\cdot 10^{-3} {\rm arcseconds}/{\rm year}  \ .
\end{equation}
Assuming that $\Omega_{\phi}(\beta=1)$ is approximately $2\pi$ per day, we find that
\begin{equation}
\frac{1}{\beta}-1\lesssim 10^{-11} \Rightarrow \frac{\delta}{2\pi}\lesssim   10^{-11}  \ .
\end{equation}
This transfers to a bound on the energy per unit length $\mu$ of a cosmic string piercing the earth that reads 
\begin{equation}
\mu \lesssim 10^{16} \ {\rm kg/m} \ .
\end{equation}
Surprisingly, this upper bound is in very good agreement with that found for a cosmic string piercing
a Schwarzschild black hole when comparing the theoretical results with the experimental results for the perihelion shift of Mercury and the light deflection by the Sun \cite{hhls}.

\section{Conclusions}
It would be fascinating to detect cosmic strings in the universe since this detection
would open the window to the physics of the very early universe and could prove (or disprove)
theories such as string theory, supersymmetry and Grand Unification. In most cases the prediction
of the detection of these objects has focused on the Cosmic Microwave background (CMB) data \cite{cmb_cosmic}.
Here we discuss another possibility, namely that cosmic strings might be detected due to
the way that test particles move in their space--time. Hence,  
in this paper we have studied the analytical solutions of the geodesic
equation in the space--time of a Kerr black hole pierced by an infinitely thin cosmic string aligned with the rotation axis of the black hole. We have given the analytical
solutions for the $t$-, $r$-, $\theta$- and $\phi$-components of the geodesic equation in terms of
Mino time which allows to decouple the $r$- and $\theta$-motion. We see that the main difference
to the standard Kerr space--time (which corresponds to the limit of vanishing deficit angle) is the 
change of the $\phi$-motion. In particular the precession of the orbital plane as well
as of the orbital ellipse of bound orbits will increase for increasing deficit angle, i.e. increasing
energy per unit length of the string. Comparing our results with the LAGEOS
data, we find that the upper limit for the energy per unit length of a cosmic string piercing the earth is $\mu \lesssim 10^{16}$ kg/m. Our results have also applications in the computation
of gravitational wave templates for extreme mass ratio inspirals and to the recently suggested alignment of
the polarization vector of quasars, respectively. 
\\
\\
{\bf Acknowledgments}
The work of PS was supported by DFG grant HA-4426/5-1.

\section{Appendix}

\subsection{Domain of existence for the $\theta$--motion}
In order to find the expressions in (\ref{lzmassive}) and (\ref{lzmassless}) we use the variable
$v:=\cos^2\theta$ and rewrite (\ref{thetalambda}) as follows
\begin{equation}
 d\lambda=-\frac{dv}{\sqrt{\Theta_v}}
\end{equation}
with $\Theta_v=-4a^2(E^2+\varepsilon)v^3 +4\left(a^2(E^2+\varepsilon)-L_z^2 -{\cal{Q}}\right)v^2 +4{\cal{Q}}v$.
For $v \in ]0:1[$ we assume that two of the zeros merge, i.e. we can then write $\Theta_v$ as follows
\begin{equation}
 \Theta_v=(v-\xi_1)^2\xi_2v \,
\end{equation}
where $\xi_1$ and $\xi_2$ can be given as functions of $L_z$, $E$, $\varepsilon$ and $\cal{Q}$.  These relations
can then be used to give $L_z$ in terms of $E$.

\subsection{Computation of $I_{\theta}$}
Using the substitutions of section \ref{thetasection} the expression $dI_{\theta}$ reads~:
\begin{eqnarray}
dI_{\theta}=-\frac{L_{z}dv}{(1-v)\sqrt{(\Theta_{v})}}+\frac{aE dv}{\sqrt{\Theta_{v}}}=\frac{\mu(L_z-aE)dw}{\sqrt{4w^3-g_{2}w-g_3}}+\frac{L_z (\nu+\mu d_{0})dw}{(w-d_0)\sqrt{4w^3-g_{2}w-g_3}}\label{Itheta3}  \ ,
\end{eqnarray}
where
$d_0=\frac{1-\nu}{\mu}$.
Let us define the new variable $x$ and the constant $x_0$ by using the Weierstrass $\wp$ function~:
\begin{eqnarray}
x:=\int_{\infty}^{w}\frac{dw}{\sqrt{4w^3-g_2w-g_3}}\quad\mbox{and}\quad x_0:=\int_{\infty}^{w_0}\frac{dw}{\sqrt{4w^3-g_2w-g_3}} \ ,
\end{eqnarray}
i.e. $w=\wp(x;g_2,g_3)$.
With this definition $x$ can be solved as a function of Mino time $\lambda$~:
\begin{eqnarray}
x-x_{0}=\frac{\lambda-\lambda_{0}}{\mu} \ .
\end{eqnarray}
In addition $dw$ can be written as
\begin{eqnarray}
dw&=&\wp'(x)dx =\pm\sqrt{4\wp^3(x)-g_2\wp(x)-g_3}dx\quad\mbox{where}\quad \wp'=\frac{d\wp(x;g_2,g_3)}{dx}  \ .
\end{eqnarray}
Inserting $dw$ into (\ref{Itheta3}) gives~:
\begin{eqnarray}
dI_{\theta }&=&\mu(L_{z}-aE)dx+(\nu+\mu d_{0})\frac{dx}{\wp(x)-d_{0}}\label{Itheta2}  \ .
\end{eqnarray}
The first term in (\ref{Itheta2}) can be integrated easily. For the second term, we can use the formula~:
\begin{eqnarray}
\int_{x_{0}}^{x}\frac{dx}{\wp(x)-d_{0}}=\sum_{i=1}^{2}\frac{1}{\wp'(x_{i})}\left[(x-x_0)\zeta(x_{i})+\ln{(\sigma(x-x_i))}-\ln{(\sigma(x_0-x_i))}\right]\label{rfnWeierstrass} \ ,
\end{eqnarray}
where $\wp(x_{1})$ =$\wp(x_{2})$ = $d_{0}$. The simple poles of the function in (\ref{rfnWeierstrass}) denoted by $x_1$, $x_2$ are in the fundamental domain  $\{2a\omega_{1}+2b\omega_{2}\vert a,b\in [0,1]\}$, where $2\omega_{1}\in \mathbb{R}$ and $2\omega_{2}\in\mathbb{C}$. Hence the analytical solution of $I_{\theta}$ is~:
\begin{eqnarray}
\int dI_{\theta}=I_{\theta}&=&\mu(L_{z}-aE)(x-x_0)+\sum_{i=1}^{2}\frac{\nu+\mu d_{0}}{\wp'(x_{i})}\left[(x-x_{0})\zeta(x_{i})+\ln{(\sigma(x-x_i))}-\ln{(\sigma(x_0-x_i))}\right]\nonumber\\
&=&(L_{z}-aE)(\lambda-\lambda_{0})+\sum_{i=1}^{2}\frac{\nu+\mu d_{0}}{\wp'(x_{i})}\left[\frac{(\lambda-\lambda_{0})}{\mu}\zeta(x_{i})+\ln{(\sigma(x-x_i))}-\ln{(\sigma(x_0-x_i))}\right]\label{Ithephi}  \  \ .
\end{eqnarray}

\subsection{Computation of $I_r$}

With the substitution $r=\frac{1}{y}+ r_{1}$ the expression $dI_r$  can be written as a function of $y$ as follows~:
\begin{eqnarray}
dI_{r}&=&\frac{aP(r)dr}{\Delta\sqrt{R(r)}}
=-\left(\frac{a}{r_1^2+a^2-2Mr_1}\right)\left[\frac{(Er_1^2+Ea^2-L_za)y^2+2Er_1y+E}{y^2+\frac{2r_1-2M}{r_1^2+a^2-2Mr_1}y+\frac{1}{r_1^2+a^2-2Mr_1}}\right]\frac{dy }{\sqrt{R_y}}\label{Iryform}  \ .
\end{eqnarray}
Moreover the integral in (\ref{Iryform}) can be written in the form~:
\begin{eqnarray}
dI_{r}=\left[K_{0}+\frac{K_{1}}{y-y_{1}}+\frac{K_{2}}{y-y_{2}}\right]\frac{dy}{\sqrt{R_{y}}} \ ,
\end{eqnarray}
where
\begin{eqnarray}
y_{1,2}&=&\frac{r_1-M\pm \sqrt{M^2-a^2}}{-r_1^2-a^2+2Mr_1}
\end{eqnarray}
and the $K_{i}$, $i=0,1,2$ are constants. Using the transformation $y=\gamma z + \alpha$ (where $\gamma$ and $\alpha$ are constants) the expression $dI_r$ becomes~:
\begin{eqnarray}
dI_{r}&=&K_0\frac{\gamma dz}{\sqrt{4z^{3}-\tilde{g}_2z-\tilde{g}_3}}+\sum_{j=1}^{2}K_{j}\frac{dz}{(z-e_{j})\sqrt{4z^{3}-\tilde{g}_2z-\tilde{g}_3}}\quad\mbox{where}\quad e_{j}:=\frac{y_j-\alpha}{\gamma}\label{Irzform}
\end{eqnarray}
We introduce the new variable $u$ as follows~:
\begin{eqnarray}
u :=\int_{\infty}^{z}\frac{dz}{\sqrt{4z^{3}-\tilde{g}_2z-\tilde{g}_3}} \ ,
\end{eqnarray}
i.e. $\wp(u;\tilde{g}_2,\tilde{g}_3)=z$.
Then $u$ can be given as a function of Mino time $\lambda$. Finally using
(\ref{Itheta2})-(\ref{rfnWeierstrass}) the analytical solution of $I_{r}$ is~:
\begin{eqnarray}
\int dI_r=I_{r}&=&\gamma K_{0}(u-u_{0})+\sum_{i,j=1}^2\frac{ K_{j}}{\wp'(u_{ji})}\left[(u-u_{0})\zeta(u_{ji})+\ln{(\sigma(u-u_{ji}))}-\ln{(\sigma(u_{0}-u_{ji}))}\right]\nonumber\\
&=&-K_{0}(\lambda-\lambda_{0})+\sum_{i,j=1}^2\frac{ K_{j}}{\wp'(u_{ji})}\left[\frac{-(\lambda-\lambda_{0})}{\gamma}\zeta(u_{ji})+\ln{(\sigma(u-u_{ji}))}-\ln{(\sigma(u_{0}-u_{ji}))}\right]\label{Irphi}  \ ,
\end{eqnarray}
where
$u=-\left(\frac{1}{\gamma}(\lambda-\lambda_{0})+C_r\right)$, $u_{0}=-C_r$
and $u_{ji}$ =$e_j$ with $\wp(u_{11})$ = $\wp(u_{12})$ = $e_1$ and $\wp(u_{21})$ = $\wp(u_{22})=e_2$.

\end{document}